\documentclass[showpacs,showkeys,11pt,
preprint,preprintnumbers,nofootinbib,
groupedaddress,superscriptaddress]{revtex4-1}

\usepackage[dvipdfmx]{graphicx}
\usepackage{cancel,subfigure}
\usepackage{color}
\usepackage{amsmath,amssymb}	
\usepackage{multirow}
\usepackage{enumerate}

\allowdisplaybreaks

\begin{document}
\title{Diphoton resonance at the ILC}
\preprint{}
\author{Junping Tian}
\email{tian@icepp.s.u-tokyo.ac.jp}
\affiliation{International Center for Elementary Particle Physics, The
University of Tokyo, Tokyo 113-0033, Japan}
\author{Keisuke Fujii}
\email{keisuke.fujii@kek.jp}
\affiliation{KEK, Tsukuba 305-0801, Japan}
\author{Hiroshi Yokoya}
\email{hyokoya@kias.re.kr}
\affiliation{Quantum Universe Center, KIAS, Seoul 130-722, Republic of
Korea}


\date{\today}

\begin{abstract}
In this paper we study the direct production of the diphoton resonance
 $X$ which has been suggested by 2015 data at the LHC, in $e^+e^-\to
 X\gamma/XZ$ processes at the ILC. 
We derive an analytic expression for the scattering amplitudes of these
 processes, and present a comprehensive analysis for determining the
 properties of $X$ at the ILC.
A realistic simulation study for $e^+e^-\to X\gamma$ is performed based
 on the full detector simulation to demonstrate the capabilities of the
 ILC experiment.
Complementary to the searches at the LHC, prospects of the measurement
 of the absolute values of production cross-section are obtained for the ILC using
 recoil technique without assuming decay modes of $X$.
In addition, we have studied the searches for $X\to\rm{invisible}$ and
 $X\to b\bar{b}$ modes, which are challenging at the LHC, 
 and found that these decay modes can be discovered with high significance
 if their branching ratios are large enough.
\end{abstract}
\maketitle

\section{Introduction}

At the end of 2015, both the ATLAS and CMS Collaborations have reported
an excess in the diphoton mass spectrum near
$m_{\gamma\gamma}\simeq750$~GeV by using the LHC data collected at
$\sqrt{s}=13$~TeV~\cite{Aaboud:2016tru,Khachatryan:2016hje}.
Although the statistical significance of their findings has not been
conclusive yet - local (global) significances by ATLAS~\cite{Aaboud:2016tru,Aad:2015mna} and CMS~\cite{Khachatryan:2016hje,Khachatryan:2015qba} are 3.9$\sigma$ (2.1$\sigma$) 
and 3.4$\sigma$ (1.6$\sigma$), respectively - it has
motivated many particle physicists to propose models for the physics
beyond the standard model~(SM), and to study their phenomenological
implications to reveal the physics behind the observed
excess~\cite{Franceschini:2015kwy,Strumia:2016wys,Franceschini:2016gxv}.
As there have appeared various scenarios to explain the observed excess,
what we could conclude so far about the observed excess is limited.

In this paper, we study the prospect of investigating the physics
behind the observed excess, if it is indeed coming from some new physics, 
at the future International Linear
Collider~(ILC) experiment~\cite{Baer:2013cma,Fujii:2015jha}.
We restrict ourselves to a ``standard'' scenario that there
exists a new particle $X$ having a mass around 750~GeV and a
coupling to a pair of photons, and study the experimental methods for
profiling $X$ taking advantage of clean environments of electron-positron
collider experiments~\cite{Djouadi:2016eyy,Richard:2016nhm,Ito:2016kvw,Bae:2016oey}.
We also focus our studies on the $e^+e^-$ collision instead of
$\gamma\gamma$ collision~\cite{Djouadi:2016eyy,Richard:2016nhm,Ito:2016zkz,He:2016olo},
and on the associate production: $e^+e^-\to X\gamma$ or $XZ$
despite the possibility of $s-$channel production through $e^+e^-\to X$ if $X$
couples to $e^+e^-$ directly~\cite{Giddings:2016sfr,Richard:2016nhm}.
Comparing to the existing feasibility studies of $X$ resonance at lepton
colliders~\cite{Djouadi:2016eyy,Ito:2016kvw,Bae:2016oey}, the new aspects 
of this paper are on the following two points.
Firstly, we present comprehensive theoretical analysis on the
$X\gamma/XZ$ productions at the ILC taking into account the
beam polarisations and angular distributions, which are useful for CP
measurement.
Secondly, we give results based on
 full detector simulation and realistic beam beam interactions.
Complementary to the search at the LHC, we explore the measurement of 
absolute production cross sections as well as the partial decay
widths of $X$ using recoil technique, which are possible only at lepton
colliders.
Furthermore, we study the searches for $X\to\rm{invisible}$ decay, which is
interesting in the models where $X$ is connected to dark
matter, and for $X\to b\bar{b}$ decay, which is very challenging to look for at
the LHC.

It is worth emphasising that even though the studies are done for the
unconfirmed $X$ resonance, the theoretical calculations and experimental
methods developed in this paper are quite general and useful for studies
at lepton colliders of any similar new particles which couple to diphoton.
This paper can hence be considered as a case study for new
particle search at the ILC. 
The paper is organised as follows.
We introduce a theoretical framework in Sec.~\ref{Sec:theory}, and 
present analytic results for $e^+e^-\to X\gamma/XZ$ in
Sec.~\ref{Sec:XatILC}.
Section~\ref{Sec:simulation} is devoted to realistic simulation
studies at the ILC.
We present the summary and conclusions of this paper in
Sec.~\ref{Sec:summary}.

\section{Theoretical framework} 
\label{Sec:theory}
The coupling of a neutral boson to a pair of photons is naturally
interpreted by loop diagrams of (the SM or new) particles which are
charged under the  SU(2)$_L$ and/or U(1)$_Y$ gauge groups of
SM. 
In general, these also induce couplings to the other pairs of the
bosons, such as $Z\gamma$, $ZZ$, and $W^+W^-$. 
In addition, if the internal particle is charged under SU(3)$_c$, a
coupling to a pair of gluons is induced as well. 
Without specifying the details of the model for $X$, but keeping
generality of the nature of $X$, we consider the effective couplings of
$X$ to the SM gauge bosons based on the effective Lagrangian which is
invariant under the SM gauge groups.
We consider $X$ as a spin-0 parity-even or odd particle.
However, in this study, we do not consider $X$ as a spin-2 particle,
since the realisation of the model for a massive spin-2 particle
compatible with the observed diphoton excess and the other LHC constraints
requires details of the theory configuration, and the collider
signatures at the ILC as well as at the LHC must depend on the details.
See, e.g.\ Refs~\cite{Arun:2015ubr,Han:2015cty,Martini:2016ahj,%
Geng:2016xin,Sanz:2016auj,Bernon:2016dow,He:2016olo,Giddings:2016sfr}
for studies of the spin-2 scenario. 

In the case where $X$ is a scalar (spin-0 parity-even), we define the
effective Lagrangian, 
\begin{align}
 {\mathcal L}_{X_S} = -\frac{1}{4\Lambda}\left[
 c_1 B_{\mu\nu}B^{\mu\nu} + c_2 W^k_{\mu\nu}W^{k\,\mu\nu} + c_3
 G^a_{\mu\nu}G^{a\,\mu\nu} 
 \right] X_S, \label{eq:lag0}
\end{align}
where the field strength is defined as $F_{\mu\nu}=\partial_\mu
F_\nu-\partial_\nu F_\mu + \cdots$ for $F=B,W^k,G^a$, the gauge fields
for U(1)$_Y$, SU(2)$_L$, and SU(3)$_{c}$, respectively. 
Summations for $k=1,2,3$ and $a=1,\cdots,8$ are implicit. 
For simplicity, we have neglected the cubic and quadratic terms of
the SU(2)$_L$ and SU(3)$_{c}$ gauge fields since these are irrelevant in
this paper. 
We introduce a common cut-off scale $\Lambda$ and coupling constants
$c_i$ for $i=1$, 2, 3.
In terms of the field strengths for physical states,
$A_\mu=c_wB_\mu+s_wW^3_\mu$, $Z_\mu=-s_wB_\mu+c_wW^3_\mu$, 
$W^\pm_\mu=(W^1_\mu \pm iW^2_\mu)/\sqrt{2}$, 
where the weak mixing angle is defined as $c_w=\cos\theta_w$,
$s_w=\sin\theta_w$ with $s_w^2=1-M_W^2/M_Z^2$, the Lagrangian can be re-written
as
\begin{align}
 {\mathcal L}_{X_S} = -\frac{1}{4\Lambda}\left[
 c_\gamma A_{\mu\nu}A^{\mu\nu}
 + c_{\gamma Z}A_{\mu\nu}Z^{\mu\nu}
 + c_{Z}Z_{\mu\nu}Z^{\mu\nu}
 + c_{W}W^{+}_{\mu\nu}W^{-\,\mu\nu}
 + c_gG^a_{\mu\nu}G^{a\,\mu\nu}
 \right] X_S. \label{eq:lag0b}
\end{align}
The effective couplings for the physical states are given in terms of
$c_i$ and weak mixing angles as 
\begin{align}
 & c_\gamma     = c_w^2c_1 + s_w^2c_2,  \\
 & c_{\gamma Z} = -2c_ws_w (c_1 - c_2), \\
 & c_Z          = s_w^2c_1 + c_w^2c_2,  \\
 & c_W          = 2c_2, \\
 & c_g          = c_3.
\end{align}
The partial decay widths of $X_S$ into gauge bosons are calculated in
the Appendix. 

For the pseudoscalar (spin-0 party-odd) case, the effective Lagrangian
is given as, 
\begin{align}
 {\mathcal L}_{X_P} &= -\frac{1}{4\Lambda}\left[
 \tilde c_1 B_{\mu\nu}\widetilde B^{\mu\nu}
 + \tilde c_2 W^a_{\mu\nu}\widetilde W^{a\,\mu\nu} 
 + \tilde c_3 G^a_{\mu\nu}\widetilde G^{a\,\mu\nu}\right]
 \label{eq:lagP} \\
 & = 
 -\frac{1}{4\Lambda}\left[
 \tilde c_\gamma     A_{\mu\nu}\widetilde{A}^{\mu\nu}
 + \tilde c_{\gamma Z} A_{\mu\nu}\widetilde{Z}^{\mu\nu}
 + \tilde c_Z          Z_{\mu\nu}\widetilde{Z}^{\mu\nu}
 + \tilde c_W         W^+_{\mu\nu}\widetilde{W}^{-\,\mu\nu} 
 + \tilde c_g      G^a_{\mu\nu}\widetilde{G}^{a\,\mu\nu}
 \right] X_P, \nonumber 
\end{align}
where $\widetilde A^{\mu\nu}$ is defined as $\widetilde
A^{\mu\nu}=1/2\cdot\epsilon^{\mu\nu\alpha\beta}A_{\alpha\beta}$.
Similarly to the scalar case, the couplings $\tilde c_i$ with
$i=\gamma$, $\gamma Z$, $Z$, $W$, $g$ are given in terms of $\tilde c_i$
with $i=1$, 2, 3 in the same manner as with Eqs.~(3-7).

Within these effective Lagrangians, the production process of $X$ at the
LHC is the gluon-fusion process, $gg\to X$.
Thus the event rate of $pp\to X\to\gamma\gamma$ is proportional to
$\Gamma_{gg}\cdot\Gamma_{\gamma\gamma}/\Gamma_X$ where $\Gamma_X$ is the
total decay width of $X$.
$\Gamma_{gg}$ has to satisfy the constraint from the dijet resonance
searches at the LHC, $\Gamma_{gg}/M_X\lesssim
10^{-3}$~\cite{CMS:2015neg}.
By making assumptions on $\Gamma_X$, the values of $\Gamma_{gg}$ and
$\Gamma_{\gamma\gamma}$ are bounded by the size of the observed excess
in the current data.
We consider two benchmark scenarios (common for both scalar and
pseudoscalar cases): one assuming a large decay width
suggested by the current ATLAS analysis~(BP1), and the other assuming
a minimum set of the decay modes, $\Gamma_X \simeq \Gamma_{gg} + 
\Gamma_{\gamma\gamma}$~(BP2).
As a typical value of $\Gamma_{\gamma\gamma}$, we obtain
$\Gamma_{\gamma\gamma}/M_X=10^{-3}$~($10^{-5}$) in BP1~(BP2).
Any value of $\Gamma_X$ between the two benchmark points or even larger one
 can be assumed without conflicting current data.
A short summary of the benchmark points is presented in Table~\ref{tab:BP}.

\begin{table}[b]
 \begin{tabular}{c ccccc}
  & $\Gamma_{gg}            /M_X$ &
   $\Gamma_{\gamma\gamma}  /M_X$ &
  $\Gamma_{\rm tot}       /M_X$ &
   ${\rm Br}_{gg}          $ &
  ${\rm Br}_{\gamma\gamma}$ \\
  \hline \hline 
  BP1 & $10^{-5}$ & $10^{-3}$ & 0.06 & 0.017\% & 1.67\% \\
  BP2 & $10^{-6}$ & $10^{-5}$ & $1.1\times10^{-5}$ & 9.1\% & 90.9\% \\
  \hline
 \end{tabular}
 \caption{Summary of the two benchmark points, BP1 and
 BP2.} \label{tab:BP}
\end{table}

In BP1, the rest of the decay modes can be those to any other SM
particles, such as the other pairs of the SM gauge bosons depending on
the parameters in the effective Lagrangian given above, $\ell^+\ell^-$,
$jj$, $b\bar b$, and $t\bar t$ within the constraints by direct searches
of the resonance in these decay modes.
Alternatively, it could be dominated by decays into invisible particles such as
neutrinos or dark matter, which are poorly constrained at the LHC.
Constraints on the branching ratios of $X\to Z\gamma$, $ZZ$, and $W^+W^-$
by the LHC 8~TeV data put upper limits on the ratios of
branching ratios to $X\to\gamma\gamma$
as~\cite{Franceschini:2015kwy,Alves:2015jgx}
\begin{align}
 \frac{\Gamma_{Z\gamma}}{\Gamma_{\gamma\gamma}}\lesssim 2,\quad
 \frac{\Gamma_{ZZ}}{\Gamma_{\gamma\gamma}}\lesssim 6,\quad
 \frac{\Gamma_{WW}}{\Gamma_{\gamma\gamma}}\lesssim 20.
\end{align}
It has been pointed out that the $Z\gamma$ mode gives the most stringent
constraint on $r\equiv{c_2}/{c_1}$; only the region $-0.6\lesssim
r\lesssim 6.4$ is allowed by the LHC Run-I data for both scalar and
pseudoscalar cases.

\section{$X(750)$ production at the ILC}
\label{Sec:XatILC}
In $e^+e^-$ collisions, $X$ can be produced via the following
processes~\cite{Djouadi:2016eyy,Ito:2016kvw}:
\begin{align}
 & e^+ e^- \to \gamma^*/Z^* \to X \gamma,\label{eq:XA}\\
 & e^+ e^- \to \gamma^*/Z^* \to X Z,\label{eq:XZ}\\
 & e^+ e^- \to e^+ e^- X\quad(Z/\gamma~{\rm fusion}),\\
 & e^+ e^- \to \nu_e\bar\nu_e X\quad(W~{\rm fusion}).
\end{align}
In addition, $\gamma\gamma\to X$ production with the photon-photon
collider option at the future lepton colliders has been examined in
Ref.~\cite{Ito:2016zkz,He:2016olo}.

We study processes (\ref{eq:XA}) and (\ref{eq:XZ}) with the ILC
experiment at the center-of-mass energy larger than 750~GeV.
In the measurement of these processes at lepton colliders, $X$ can be
identified without looking at its decay products.
It can be identified as a peak in the recoil mass distribution in the
inclusive production of a hard photon or $Z$.
By the searches using the recoil mass technique, the resonance can be
identified even if it decays dominantly into invisible final-states.

\subsection{$e^+e^-\to X\gamma$}

First, we calculate the scattering amplitudes for process 
(\ref{eq:XA}) analytically, to evaluate the total cross-section as
well as the differential distributions. 
Four-momentum and helicity of each particle are assigned as follows:
\begin{align}
 e^- (k_1,\sigma_1) + e^+ (k_2,\sigma_2) \to X_{0} (p_1) + \gamma
 (p_2,\lambda_\gamma) .
\end{align}
The scattering amplitudes are calculated to be
\begin{align}
 {\mathcal M}_{X_S\gamma}(\lambda_V,\lambda_\gamma) =
 \frac{e}{\Lambda}\sqrt{\frac{s}{2}} \beta
 \lambda_V\frac{1-\lambda_V\lambda_\gamma\cos\theta}{2}
{\mathcal A}_{\gamma}(s;\lambda_V),
\end{align}
where we define $\beta=1-M_X^2/s$,
\begin{align}
 {\mathcal A}_{\gamma}(s;\lambda_V) = c_\gamma
 -\frac{c_{\gamma Z}}{2} \frac{c_V^{e}-\lambda_V c_A^e}{2c_ws_w}r_Z(s),
\end{align}
and $r_Z(s)=1/(1-M_Z^2/s)$.
The vector and axial-vector couplings of electron are
$c_V^e=-\frac{1}{2}+2s_w^2$ and $c_A^e=-\frac{1}{2}$, respectively.
$s=2k_1\cdot k_2=2p_1\cdot p_2+M_X^2$ is the square of the total
collision energy in the $e^+e^-$ center-of-mass frame, and
$\theta$ is the scattering angle in the laboratory frame.
Without loss of generality, we set the azimuthal angle to zero,
$\phi=0$.
$\lambda_V\equiv\sigma_1-\sigma_2$ is the difference of the
helicities of the beam electron and positron, and $\lambda_\gamma$ is
the photon helicity.
$\lambda_V$ can be $\pm1$ and 0, while $\lambda_\gamma=\pm1$.
For $\lambda_V=0$, i.e.\ $\sigma_1=\sigma_2$, the amplitude is zero.
Summing over the helicity of the final-state photon, the scattering
angle distribution is obtained to be
\begin{align}
 \frac{d\sigma_{X_S\gamma}}{d\cos\theta} = \,
 & \frac{\alpha\beta^3}{16\Lambda^2} \frac{1+\cos^2\theta}{2}
 \left|{\mathcal A}_{\gamma}(s;\lambda_V)\right|^2,
\end{align}
for $\lambda_V=\pm1$, where $\alpha$ is a fine-structure constant
$\alpha=e^2/(4\pi)$. 
Finally, the total cross section is calculated to be 
\begin{align}
\sigma_{X_S\gamma} \, &=
 \frac{\alpha\beta^3}{12\Lambda^2} 
 \left|{\mathcal A}_{\gamma}(s;\lambda_V)\right|^2,
\end{align}
for $\lambda_V=\pm1$.
Similarly, the scattering amplitudes for the pseudoscalar case is
calculated to be
\begin{align}
 {\mathcal M}_{X_P\gamma}(\lambda_V,\lambda_\gamma)
 = i\frac{e}{\Lambda} \sqrt{\frac{s}{2}}\beta
 \lambda_V\lambda_\gamma\frac{1-\lambda_V\lambda_\gamma\cos\theta}{2}
\tilde{\mathcal A}_{\gamma}(s;\lambda_V) ,
\end{align}
where $\tilde{\mathcal A}_\gamma$ is defined similarly to ${\mathcal A}_\gamma$
by replacing $c_i$ to $\tilde{c}_i$.
Since the structure of the scattering amplitudes is completely the same
as that for the scalar case at the Born level, the differential
distribution as well as the total cross section for the pseudoscalar
case is obtained from those for the scalar case by replacing the
coupling constants $c_i\to \tilde c_i$. 
The values of the coupling constants obtained by fitting the LHC
diphoton resonance excess are also same for the scalar and pseudoscalar 
cases, thus we find no observable which can distinguish the parity of
the resonance in $e^+e^-\to X\gamma$ process.

\begin{figure}[t]
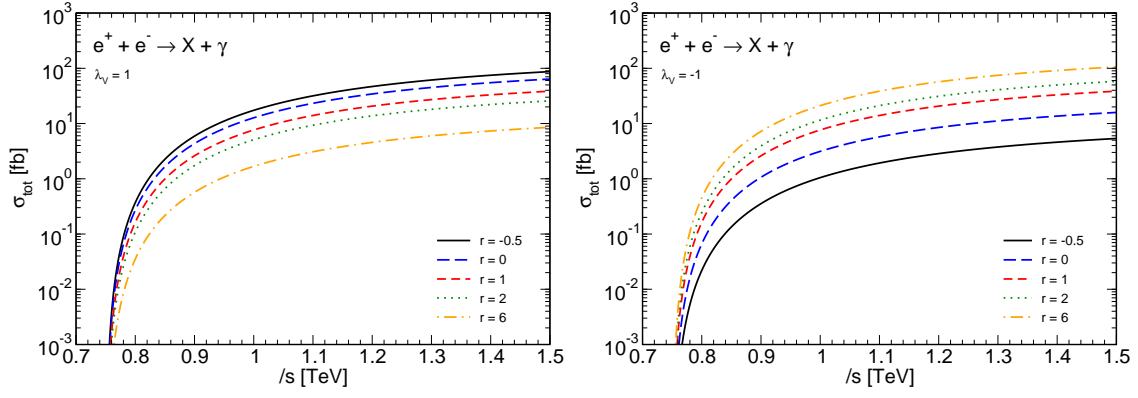

 \includegraphics[width=.45\textwidth]{./sigtotXA.eps}
 \includegraphics[width=.45\textwidth]{./sigtotXA2.eps}
 \caption{Total cross section of $e^+e^-\to X\gamma$ for
 $r(=c_2/c_1)=-0.5$, 0, 1, 2, 6, as a function of the collision
 energy. Left: $\lambda_V=1$, right: $\lambda_V=-1$. }\label{fig:sig0}
\end{figure}

\begin{figure}[t]
 \begin{center}
  \includegraphics[width=.5\textwidth]{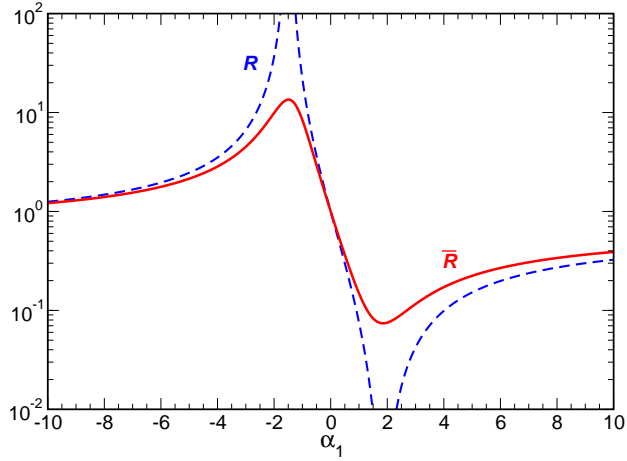}
  \caption{Ratios of the polarised cross sections as a function of
  $\alpha$; ${\mathcal R}$: purely polarised beams ($\lambda_V=\pm1$);
  $\overline{\mathcal R}$:
  $(P_{e^-},P_{e^+})=(\pm0.8,\mp0.2)$.}\label{fig:r1}
 \end{center}
\end{figure}

In Fig.~\ref{fig:sig0}, we show the total cross section of $e^+e^-\to
X\gamma$ for BP1 as a function of the collision energy.
The cross section for BP2 is typically two orders of magnitude smaller
than that for BP1. 
For $\lambda_V=1$ the cross section is larger for smaller $r$, while
for $\lambda_V=-1$ the cross section is larger for larger $r$.
Thus, the effects of $Z$-mediated diagram can be clearly observed in the
production rates with the polarized beams.
In the case of $r=0$ where $X$ couples to the U(1)$_Y$ gauge field
only, the cross section is proportional to the square of the hypercharge
of electrons.
Therefore, $\sigma_{X\gamma}(\lambda_V=1) /
\sigma_{X\gamma}(\lambda_V=-1) = (Y_{e_R}/Y_{e_L})^2 = 1/4$. 
We note that the ratio of the cross section can be determined by
experimental measurements without knowing the branching ratios of $X$,
and the overall strength of the couplings.
On the bases with physical boson states, the ratio is given as a
function of $\alpha_1\equiv c_{\gamma Z}/(2c_\gamma)$,
\begin{align}
 {\mathcal R}_{\gamma}(\alpha_1) =
 \frac{\sigma_{X\gamma}(\lambda_V=+1)}
 {\sigma_{X\gamma}(\lambda_V=-1)} =
 \frac{ \left|
 1 - \alpha_1\,\bar c_L^e r_Z(s) \right|^2 }{ \left|
 1 - \alpha_1\,\bar c_R^e r_Z(s) \right|^2 },
 \label{eq:alpha}
\end{align}
where $\bar c_{L/R}^e=(c_V^e\pm c_A^e)/(2c_ws_w)$.
For the realistic situation of the beam polarisation at the ILC, the
ratio of the production rates for $(P_{e^-},P_{e^+})=(0.8,-0.2)$ to 
$(P_{e^-},P_{e^+})=(-0.8,0.2)$ is given by
\begin{align}
 \overline{\mathcal R}_{\gamma}(\alpha_1) =
  \frac{\sigma_{X\gamma}[(P_{e^-},P_{e^+})=(0.8,-0.2)]}
 {\sigma_{X\gamma}[(P_{e^-},P_{e^+})=(-0.8,0.2)]}
 = \frac{{\mathcal R}_{\gamma}(\alpha_1) + \epsilon}
 {1+\epsilon{\mathcal R}_{\gamma}(\alpha_1)}, 
\label{eq:alphaB}
\end{align}
where $\epsilon=0.1\cdot0.4/(0.9\cdot0.6)\simeq0.074$.
Thus, the ratio ${\mathcal R}_{\gamma}$($\overline{\mathcal
R}_{\gamma}$) is a good probe of the $\gamma ZX$ interaction.
For $\sqrt{s}=1$~TeV, $\overline{\mathcal R}_{\gamma} \simeq
(1.-0.91\alpha_1+0.30\alpha_1^2) /
(1.+1.18\alpha_1+0.44\alpha_1^2)$. 
By solving Eq.~(\ref{eq:alphaB}), $\alpha_1$ can be determined (up to a
two-fold ambiguity), and $r=c_2/c_1$ can be further determined. 
In Fig.~\ref{fig:r1}, we plot ${\mathcal R}_{\gamma}$ and
$\overline{\mathcal R}_{\gamma}$ as a function of $\alpha_1$ for
$-10<\alpha_1<10$. 
%

\subsection{$e^+e^-\to XZ$}

Next, we consider the associate production of $X$ with a $Z$-boson at
$e^+e^-$ colliders.
With the $\gamma ZX$ and $ZZX$ vertexes introduced in the previous
section, the process occurs at the Born-level through the $s$-channel
diagrams:
\begin{align}
 e^- (k_1,\sigma_1) + e^+ (k_2,\sigma_2) \to X_{0} (p_1) + Z
 (p_2,\lambda_Z) . 
\end{align}
Scattering amplitudes for the scalar case are calculated to be
\begin{align}
 {\mathcal M}_{X_SZ}(\lambda_V,\lambda_Z=\pm1) =&\,
 \frac{e}{\Lambda}
 \sqrt{\frac{s}{2}}\sqrt{\beta_Z^2+\frac{4M_Z^2}{s}}
 \lambda_V \frac{1-\lambda_V\lambda_Z\cos\theta}{2}
 {\mathcal A}_{Z}(s;\lambda_V)
 \label{eq:ampSZ} , \\
 {\mathcal M}_{X_SZ}(\lambda_V,\lambda_Z=0) =&\,
 \frac{e}{\Lambda} M_Z\cdot\sin\theta\cdot
 {\mathcal A}_{Z}(s;\lambda_V)
 , \label{eq:ampSZ0}
\end{align}
and for the pseudoscalar case,
\begin{align}
 {\mathcal M}_{X_PZ}(\lambda_V,\lambda_Z) =&\,
 i\frac{e}{\Lambda}
 \sqrt{\frac{s}{2}}\beta_Z
 \lambda_V\lambda_Z\frac{1-\lambda_V\lambda_Z\cos\theta}{2}
\tilde{\mathcal A}_{Z}(s;\lambda_V) , \label{eq:ampPZ}
\end{align}
where
\begin{align}
 {\mathcal A}_{Z}(s;\lambda_V) = \frac{c_{\gamma Z}}{2}
 - c_Z \frac{c_V^{e}-\lambda_V c_A^e}{2c_ws_w}r_Z(s),
\end{align}
and $\tilde{\mathcal A}_{Z}$ is given by the same formulae as for ${\mathcal A}_{Z}$ after
replacing $c_i$ by $\tilde{c}_i$.
$\beta_Z = \lambda(1,M_X^2/s,M_Z^2/s)$ with
$\lambda(a,b,c)=\sqrt{a^2+b^2+c^2-2(ab+bc+ca)}$.
Note that the amplitudes for the longitudinal $Z$-boson production
vanish for the pseudoscalar case.
For the scalar case, the ratio of the longitudinal to
transverse $Z$-boson production amplitudes is proportional to $M_Z/\sqrt{s}$.

\begin{figure}[t]
 \begin{center}
  \includegraphics[width=.5\textwidth]{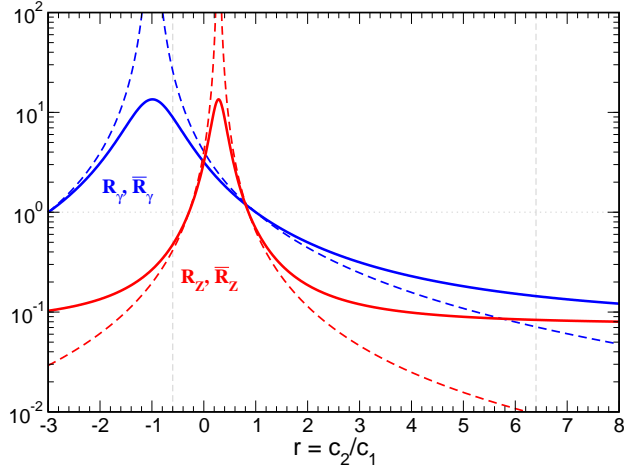}
  \caption{Ratios of the polarized cross sections as a function of
  $r=c_2/c_1$ for $X\gamma$ and $XZ$ productions.
  Blue: $X\gamma$ process, red: $XZ$ process; Dashed: ${\mathcal
  R}$, solid: $\overline{\mathcal R}$.}\label{fig:alpha} 
 \end{center}
\end{figure}

The scattering angle distributions are calculated to be
\begin{align}
 \frac{d\sigma_{X_SZ}}{d\cos\theta} =&\,
 \frac{\alpha\beta_Z}{16\Lambda^2}\left[
 \left(\beta_Z^2+\frac{4M_Z^2}{s}\right)\frac{1 + \cos^2\theta}{2}
 + \frac{2M_Z^2}{s} \sin^2\theta\right] 
\left|{\mathcal A}_{Z}(s;\lambda_V)\right|^2, \\
 \frac{d\sigma_{X_PZ}}{d\cos\theta} = &\,
 \frac{\alpha\beta_Z^3}{16\Lambda^2}
 \frac{1+\cos^2\theta}{2}\left|\tilde{\mathcal
 A}_{Z}(s;\lambda_V)\right|^2 . 
\end{align}
For the scalar case, due to the longitudinal $Z$-boson production the
scattering angle distribution has a term which behaves as
$\sin^2\theta$.
The scattering angle distributions can be written as $\propto
(1+{\mathcal B}\cos^2\theta)$, where ${\mathcal B} = \beta_Z^2 / (\beta_Z^2 +
4M_Z^2/s)$ for the scalar case, while ${\mathcal B}=1$ for the
pseudoscalar case. 
Thus, the scattering angle distributions as well
as their energy dependence can be used to distinguish the
parity of the resonance, without seeing the decay products of the
resonance.
The total cross-sections are given as
\begin{align}
 \sigma_{X_SZ} &= \frac{\alpha\beta_Z}{12\Lambda^2}
 \left(\beta_Z^2+\frac{6M_Z^2}{s}\right) \left|{\mathcal
 A}_{Z}(s;\lambda_V)\right|^2, \nonumber \\  
 \sigma_{X_PZ} &= \frac{\alpha\beta_Z^3}{12\Lambda^2}
\left|\tilde{\mathcal A}_{Z}(s;\lambda_V)\right|^2, 
\end{align}
for the scalar and psedoscalar cases, respectively.
\begin{figure}[t]
 \includegraphics[width=.45\textwidth]{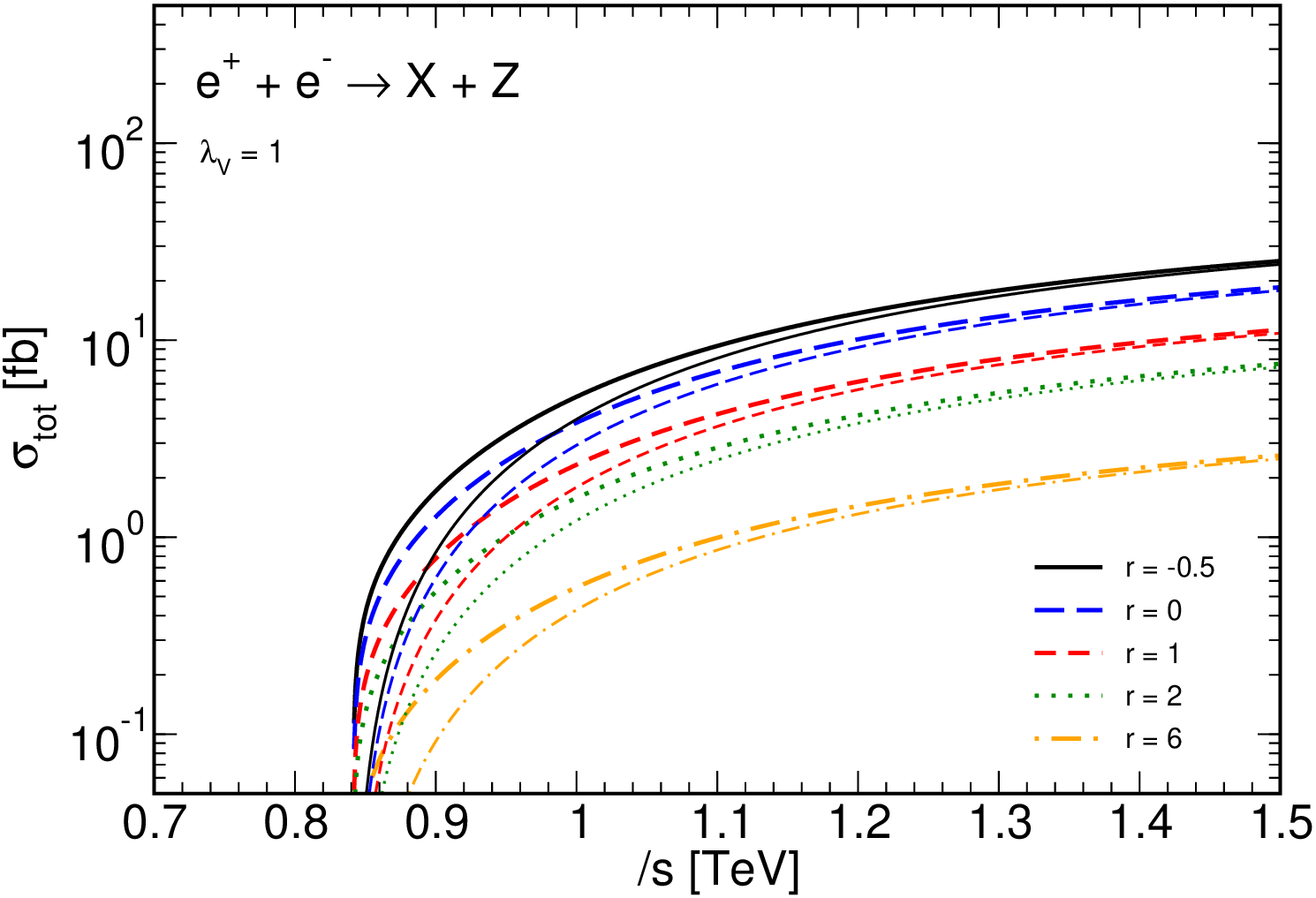}
 \includegraphics[width=.45\textwidth]{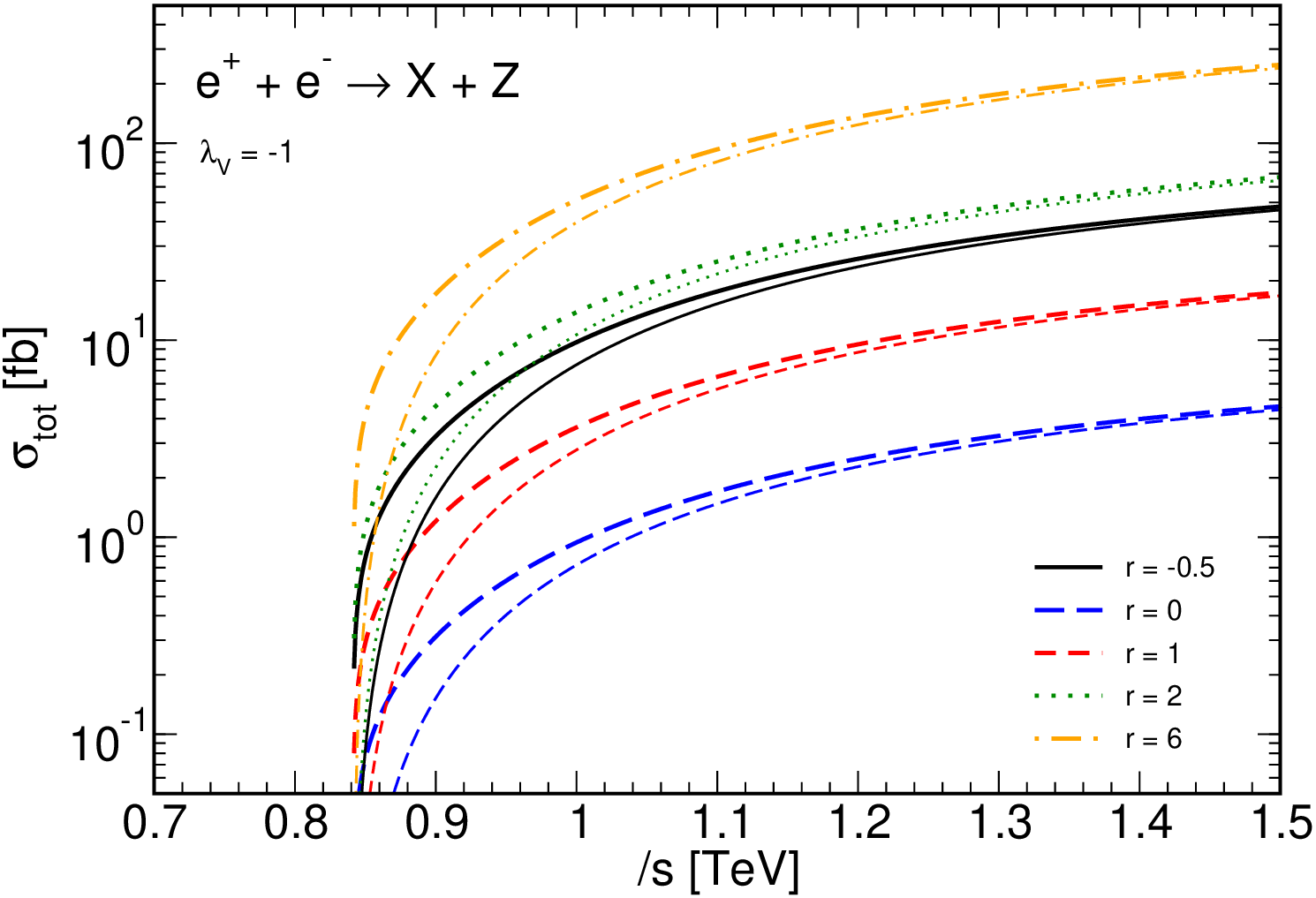}
 \caption{Total cross section of $e^+e^-\to XZ$ as a
 function of the collision energy for $r(=c_2/c_1)=-0.5$, 0, 1, 2, 6.
 Thick lines: $X_S$, thin lines: $X_P$. Left:
 $\lambda_V=1$, right: $\lambda_V=-1$.}\label{fig:sigZ0}
\end{figure}
In Fig.~\ref{fig:sigZ0}, we plot the $e^+ e^- \to XZ$ cross section
for $\lambda_V = \pm 1$ assuming the BP1 couplings.
Cross sections for the scalar and pseudoscalar cases are drawn in thick
and thin lines for $r=-0.5$, 0, 1, 2 and 6.
We find a large $r$ dependence in $\sigma_{XZ}(\lambda_V=-1)$;
those for $r=0$ and $r=6$ differ by almost two orders of magnitude. 
The ratio of the cross sections for polarized beams is given in a
similar manner to the $e^+e^-\to X\gamma$ process, with a replacement
$\alpha_1\to\alpha_2=2c_{Z}/c_{\gamma Z}$:
\begin{align}
 & {\mathcal R}_{Z} (\alpha_2) =
 \frac{\sigma_{XZ}(\lambda_V=+1)}
 {\sigma_{XZ}(\lambda_V=-1)} =
 \frac{ \left|
 1 - \alpha_2\,\bar c_L^e r_Z(s) \right|^2 }{ \left|
 1 - \alpha_2\,\bar c_R^e r_Z(s) \right|^2 }, \\
 & \overline{\mathcal R}_{Z}(\alpha_2) =
  \frac{\sigma_{XZ}[(P_{e^-},P_{e^+})=(0.8,-0.2)]}
 {\sigma_{XZ}[(P_{e^-},P_{e^+})=(-0.8,0.2)]}
 = \frac{{\mathcal R}_{Z}(\alpha_2) + \epsilon} {1+\epsilon{\mathcal
 R}_{Z}(\alpha_2)},
\end{align}
where $\epsilon=0.1\cdot0.4/(0.9\cdot0.6)\simeq0.074$.
In Fig.~\ref{fig:alpha}, we plot ${\mathcal R}_{\gamma(Z)}$ and
$\overline{\mathcal R}_{\gamma(Z)}$ as a function of $r$.
We find that for the allowed regions of $r$, $-0.6<r<6.4$, determination
of $r$ by $\overline{\mathcal R}_\gamma$ is not affected by a two-fold
ambiguity, but that by $\overline{\mathcal R}_Z$ is affected for
$r\lesssim 1$.
Determination of $r$ by the $XZ$ process would be useful for the
consistency check of the description based on the effective Lagrangian.

%

\subsection*{Lepton's angular distributions in $Z\to\ell^+\ell^-$}
Furthermore, we consider the leptonic decays of $Z$-boson, and evaluate
the angular distributions of the leptons.
In the rest frame of $Z$-boson, decay angles of $\ell^-$ is denoted as
$\hat\theta$ and $\hat\phi$, where $\hat\theta=0$ ($\hat z$-axis) gives
the direction of the $Z$-boson and $\hat\phi=0$ (and $\pi$) lies in the
scattering plane in the laboratory frame.

By defining the density matrices for the production and the decay of the final state $Z$
boson as
\begin{align}
 &\rho_{XZ}(\lambda_V;\lambda,\lambda') \propto
 {\mathcal M}_{XZ}(\lambda_V,\lambda)
 {\mathcal M}^*_{XZ}(\lambda_V,\lambda'),\\
 &\rho_{Z\to\ell^+\ell^-}(\lambda,\lambda') \propto
 {\mathcal M}_{Z\to\ell^+\ell^-}(\lambda)
 {\mathcal M}^*_{Z\to\ell^+\ell^-}(\lambda'),
\end{align}
the lepton's angular distributions are calculated to be
\begin{align}
 {\mathcal D}(\lambda_V;\theta,\hat\theta,\hat\phi) =
 \frac{3\sum_{\lambda,\lambda'}
 \left[\rho_{XZ}(\lambda_V,\lambda,\lambda')
 \cdot\rho_{Z\to\ell^-\ell^+}(\lambda,\lambda')\right]}
 {\int d\cos\theta{\rm Tr}\left[\rho_{XZ}\right]\cdot
 \int\!d\hat\Omega_2\,{\rm Tr}\left[\rho_{Z\to\ell^-\ell^+}\right]},
\end{align}
which satisfies
\begin{align}
 \int d\cos\theta d\hat\Omega_2{\mathcal
 D}(\lambda_V;\theta,\hat\theta,\hat\phi) = 1,
\end{align}
with $d\hat\Omega_2=d\cos\hat\theta d\hat\phi$.
Up to the overall normalization, the density matrix for the leptonic
decay of the $Z$-boson is obtained to be
\begin{align}
 \rho_{Z\to\ell^-\ell^+}(\lambda,\lambda') =&\,
 \left(\begin{array}{ccc}
  \frac{1+\hat c^2-2\xi\hat c}{2} & \frac{\hat s(\hat
   c-\xi)}{\sqrt{2}}e^{i\hat\phi} & \frac{\hat s^2}{2}e^{2i\hat\phi} \\
  \frac{\hat s(\hat c-\xi)}{\sqrt{2}} e^{-i\hat\phi}& \hat s^2 &
   -\frac{\hat s(\hat c+\xi)}{\sqrt{2}} e^{i\hat\phi} \\
  \frac{\hat s^2}{2}e^{-2i\hat\phi} & -\frac{\hat s(\hat
   c+\xi)}{\sqrt{2}}e^{-i\hat\phi} & \frac{1+\hat c^2+2\xi\hat c}{2}
       \end{array}\right),
\end{align}
where $\hat c=\cos\hat\theta$, $\hat s=\sin\hat\theta$, and we define
\begin{align}
 \xi = \frac{2c_V^\ell c_A^{\ell}}{(c_V^\ell)^2+(c_A^\ell)^2}.
 \label{eq:xi}
\end{align}
Integration of $\rho_Z$ over the phase-space results in
$\int
d\hat\Omega_2\rho_{Z\to\ell^-\ell^+}=8\pi/3\cdot\delta_{\lambda\lambda'}$.
The production density matrix $\rho_{XZ}$ is calculated by using the
scattering amplitudes in Eqs.~(\ref{eq:ampSZ}-\ref{eq:ampPZ}), and the
lepton's angular distributions are then calculated to be
\begin{align}
 {\mathcal D}_S(\lambda_V;\theta,\hat\theta,\hat\phi)
 & \simeq \frac{9}{128\pi}\left[(1+\cos^2\theta)(1+\cos^2\hat\theta) +
 4\lambda_V\xi\cos\theta\cos\hat\theta
 + \sin^2\theta\sin^2\hat\theta\cos2\hat\phi\right], \\
 {\mathcal D}_P(\lambda_V;\theta,\hat\theta,\hat\phi)
 & = \frac{9}{128\pi}\left[(1+\cos^2\theta)(1+\cos^2\hat\theta) +
 4\lambda_V\xi\cos\theta\cos\hat\theta
 - \sin^2\theta\sin^2\hat\theta\cos2\hat\phi\right].
\end{align}
where ${\mathcal O}(M_Z^2/s)$ terms are neglected in the scalar case.
For the reference, explicit results can be found in the Appendix.

The critical difference between the two scenarios can be seen in the
sign of the azimuthal angle dependent term,
$\pm\sin^2\theta\sin^2\hat\theta\cos2\hat\phi$. 
Integrating over $\theta$ and $\hat\theta$, and summing over the
initial-state polarisations, we get
\begin{align}
 &\widetilde{\mathcal D}_S(\hat\phi) = \frac{1}{2\pi} \left[1 +
 \frac{{\mathcal C}}{4}\cos2\hat\phi\right],\\ 
 &\widetilde{\mathcal D}_P(\hat\phi) = \frac{1}{2\pi} \left[1 -
 \frac{1}{4}\cos2\hat\phi\right], 
\end{align}
where ${\mathcal C}=(\beta_Z^2+4M_Z^2/s)/(\beta_Z^2+6M_Z^2/s)$.
Thus, observing the azimuthal angle distribution of the lepton in the
$Z$-boson decays, one can determine the spin of the resonance $X$,
without measuring its decay products.
We note that since the distribution has only $\cos2\hat\phi$ dependence,
it can be obtained without distinguishing $\ell^-$ and $\ell^+$.
It only depends on the angle between the scattering plane and the
$Z$-decay plane.
This means that the hadronic decays of $Z$-boson can be also utilised to
see this distribution, which has much larger branching fraction than the
leptonic decays.

\begin{figure}[t]
 \includegraphics[width=.5\textwidth]{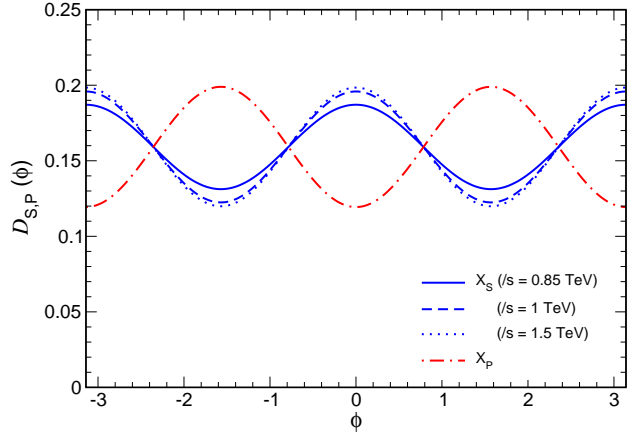}
 \caption{Azimuthal angular distribution of $Z\to\ell^+\ell^-$ in
 $e^+e^-\to XZ$ for the scalar ($S$) and pseudoscalar ($P$)
 cases.}\label{fig:PhiXZ} 
\end{figure}

In Fig.~\ref{fig:PhiXZ}, we plot the azimuthal angle distributions of
leptons for the scalar and pseudoscalar cases.
For the scalar cases, $\sqrt{s}=850$~GeV, 1~TeV and 1.5~TeV are taken,
which give ${\mathcal C}=0.7$, 0.92 and 0.99, respectively.

\subsection{Experimental determination of $X$ properties}
At the ILC, using production processes of $e^+e^-\to X\gamma$ and $XZ$, it is
possible to determine the following properties of $X$:
\begin{enumerate}[i.)]
 \item mass, which can be measured either by recoil technique
       against $\gamma$ via process (\ref{eq:XA}) or $Z$ via process
       (\ref{eq:XZ}) taking advantage of known initial state four-momentum at lepton
       colliders, or by direct reconstruction from decay particles,
       e.g.\ via $X\to\gamma\gamma, ~gg$.
 \item spin, which can be determined by looking at the angular
       distributions of $X$ production as well as its decays.
 \item CP property, which can be determined in process
       (\ref{eq:XZ}) by measuring the angular correlation of the decay
       plane of the $Z$-boson and the scattering plane spanned by the beam
       axis and the tagged $Z$-boson momentum, without measuring the $X$
       decay products.
       The distribution of the azimuthal angle ($\phi$) between the
       scattering plane and the $Z$-boson plane behaves as
       \begin{align}
	\frac{d\sigma}{d\phi}\propto 1+\frac{\mathcal C}{4}\cos2\phi,
       \end{align}
       where ${\mathcal C} = (\beta_Z^2+4M_Z^2/s)/(\beta_Z^2+6M_Z^2/s)$
       for the scalar case, while ${\mathcal C}=-1$ for the pseudoscalar
       case.
 \item total decay width ($\Gamma_X$), which can be directly measured if
       it is large (with respect to detector resolution), or can be
       indirectly determined if it is small but having sizable branching
       fractions to $\gamma\gamma$, $\gamma Z$, and $ZZ$, by using the
       recoil technique in a similar way to determining the Higgs total width, 
       $\Gamma_{X}=\Gamma(X\to VV)/{\rm Br}(X\to VV)$.
 \item values of individual effective couplings to SU(2)$_L$ and U(1)$_Y$ gauge fields, 
 which can be
       determined by measuring the dependence on beam polarisations for
       cross sections of processes (\ref{eq:XA}) and (\ref{eq:XZ}).
\end{enumerate}

\section{Event simulation and analysis}
\label{Sec:simulation}
To demonstrate the capabilities of $X$ studies at the ILC, a realistic simulation for $e^+e^-\to X\gamma$ for BP1 for spin-0 and $r=0$ has been performed with full detector simulation at $\sqrt{s} = 1$ TeV. The experimental studies for spin-2 case are expected to be quite similar. Three different modes to select $X$ are considered, one without assuming particular decay modes (inclusive mode), the other two designed specifically for $X\to\mathrm{invisible}$ and $X\to b\bar{b}$ decays. The inclusive mode is based on recoil technique, taking advantage of known initial state four-momentum, by which the $X$ invariant mass can be reconstructed using the recoil mass against $\gamma$ as $M_{\rm rec}^2 = s-2\sqrt{s}E_\gamma$ where $E_\gamma$ is the energy of reconstructed $\gamma$, henceforth $X$ can be identified without looking at its decay particles. The modes for both $X\to\mathrm{invisible}$ and $X\to b\bar{b}$ are straightforward at the ILC further by respectively requiring no significant visible energies other than that from $\gamma$ or requiring two tagged $b$-jets, thanks to the clean environment at the ILC. The searches using these three modes are complementary to those at the LHC. The inclusive mode can give the measurement of absolute production cross sections for different beam polarisations. The $X\to\mathrm{invisible}$ and $X\to b\bar{b}$ searches are very challenging at the LHC with current upper limits on partial widths of 400 and 500 times $\Gamma_{\gamma\gamma}$, respectively~\cite{Franceschini:2015kwy}. Searches using other modes such as $X\to WW,~ZZ,~\gamma\gamma,~\gamma Z,~{\rm{or}}~t\bar{t}$ are not performed in this paper because they would be measured well at the LHC if they exist. The $X\to gg$ decay is not considered in this paper because with current constraints on $\Gamma_{gg}$ and $\Gamma_{\gamma\gamma}$ the possible signal strength $\sigma(e^+e^-\to X\gamma)\times{\rm{BR}}(X\to gg)$ at the ILC can not be very large.

\begin{figure}[ht]
  \centering
 \includegraphics[width=0.9\columnwidth]{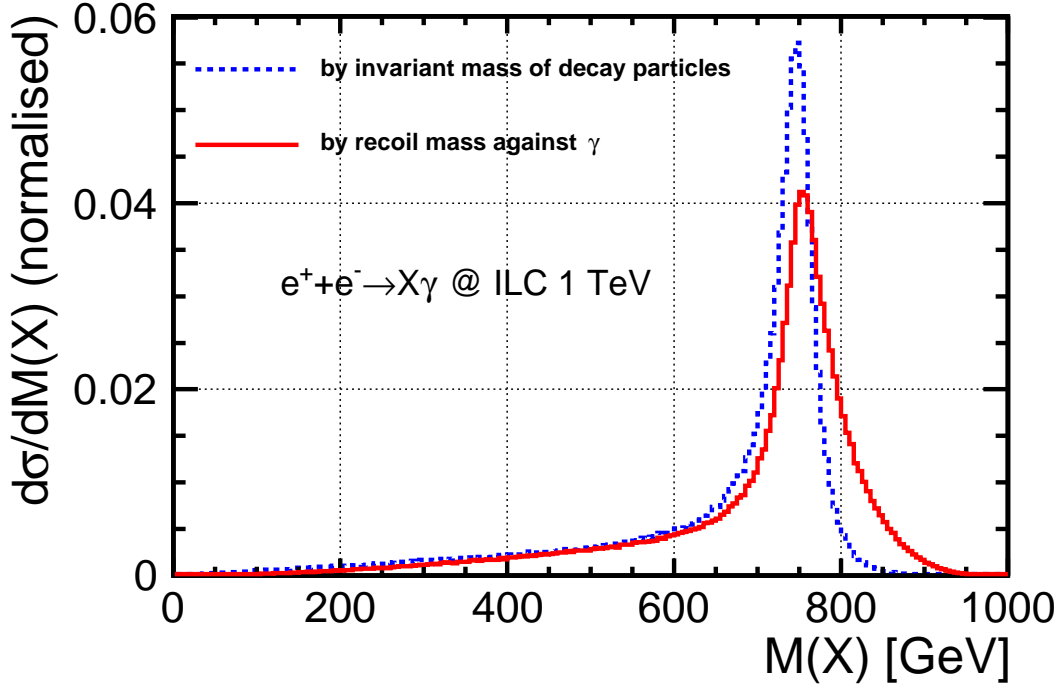}
 \caption{Mass spectrum of $X$ by recoil mass against $\gamma$ (solid, in red) and by invariant mass of decay particles (dashed, in blue) at the generator level, where large width of 45 GeV is assumed.}
 \label{fig:mrecoil_mc}
\end{figure}

It is worthwhile to point out that the realistic simulation based on full detector simulation considered in this paper is particularly important for $e^+e^-\to X\gamma$ to reliably assess the prospects at the ILC, because of the following reasons. One of the main characteristics of $e^+e^-\to X\gamma$ is the appearance of a monochromatic photon as assumed by many studies in the literatures. However, the beamstrahlung and ISR effects, which are included in this paper, would significantly modify the kinematics of final state particles. Furthermore if the total width of $X$ is as large as in BP1, implementation of the full Breit-Wigner structure for $X$ in the matrix element becomes necessary, because it will also change significantly the recoil mass spectrum, which is shown in Fig.~\ref{fig:mrecoil_mc} for
process~(\ref{eq:XA}). Regarding the background, in the inclusive mode, since the main event selection is about one photon, it is important to include all the SM background processes, any of which would survive because of the ISR effect. In the invisible search mode, it is more realistic
to include the beam induced background, such as $\gamma\gamma\to$ low
$p_t$ hadrons which will be overlaid to every event including the
signal. For the $X\to b\bar b$ search mode, it is crucial to include
full detector simulation to estimate the flavour tagging performance. 

The analysis is performed for two different beam polarisations, $P(e^-,e^+)=(-0.8,+0.2)$ (namely left-handed, or eLpR) and $P(e^-,e^+)=(+0.8,-0.2)$ (namely right-handed, or eRpL). In both cases of beam polarisations, an integrated luminosity of 2 $\rm{ab}^{-1}$ is assumed.

\subsection{Event generator and detector simulation}
The generator for $e^+e^-\to X\gamma$ is implemented using Physsim~\cite{Physsim}, where the $X\gamma\gamma$ coupling as in Eq.~(\ref{eq:lag0b}) is used, full Breit-Wigner structure of $X$ is taken into account in the matrix element \footnote{
We do not include the off-shell width effect of $X$ in the amplitudes, since the off-shell decay width of scalar to vector-vector mode 
behaves $\propto Q_X^3/M_X^3$ where $Q_X$ is the virtuality of $X$, which violates unitarity at large $Q_X$~\cite{Ko:2016xwd}. 
This violation should be canceled if our effective field theory approach is replaced with the full theory which respects unitarity. 
}, and $X$ decay into invisible or $b\bar{b}$ is considered. The generators for all background processes are obtained using Whizard 1.95~\cite{Whizard}, where all diagrams for $e^+e^-\to 2\rm{f},~4\rm{f},~\rm{or}~6\rm{f}$ (where f stands for fermion) up to parton level are included. In both Physsim and Whizard, Pythia 6.4~\cite{Pythia} is used for parton shower, fragmentation, and hadronisation. The beam spectrum including the beam energy spread and beamstrahlung is obtained by GuineaPig~\cite{GuineaPig} based on the beam parameters in TDR~\cite{ILCTDR}. The initial state radiation (ISR) spectrum for both signal and background processes is implemented using the LLA structure function~\cite{LLA}. To take into account the pile-up of beam induced background events, every signal or background event is overlaid with 4.1 events of $\gamma\gamma\to{\rm{low}}~p_t~\rm{hadrons}$ on average~\cite{TDRVol4}. The beam crossing angle of 14 mrad, which induces a common boost for every final state particle, is taken into account in the following step when the event is simulated. The total cross sections for signal and background processes are shown in Table~\ref{tab:xsec}, together with the number of expected and generated events for both left and right handed beam polarisations. It is worth mentioning that the signal cross section gets reduced by $30\%$ compared to the value given in Section 3 by analytic calculation, after beam spectrum, ISR, and Breit-Wigner structure are taken into account.

All the events from above generators are fed into a detector simulator using GEANT4~\cite{GEANT4} in the Mokka software package~\cite{Mokka} based on the ILD detector~\cite{ILD}, which is one of the two proposed detectors at the ILC. The realistic ILD design as implemented in the ILD detailed baseline design (DBD)~\cite{TDRVol4} is taken into account. The simulated events are then reconstructed in the Marlin~\cite{Marlin} framework in ILCSoft v01-16~\cite{ILCSoft}, using realistic track finding, track fitting, clustering in calorimeters, and particle flow analysis. PandoraPFA~\cite{PFA} is used for the calorimeter clustering and the particle flow analysis, which meanwhile provides photon identification for the following event selection. LCFIPlus~\cite{LCFIPlus} is used for vertex finding, jet clustering, and flavour tagging, which is relevant in this study only for the $X\to b\bar{b}$ mode. 

\begin{table}[ht]
 \begin{center}
 \begin{tabular}{| l | l | r | r | l | r | r |}
  \hline
  \multirow{2}{*}{Process}          &   \multicolumn{3}{c|}{eLpR}          &   \multicolumn{3}{c|}{eRpL}  \\
          \cline{2-7}  
          & $\sigma$ (fb) & $N_{\rm{exp}}$     &  $N_{\rm{gen}}$
          & $\sigma$ (fb) & $N_{\rm{exp}}$     &  $N_{\rm{gen}}$         \\
   \hline
   signal             &   $1.5$                      & $3.1\times 10^3$   & $2.0\times 10^5$ 
   	                 &   $4.9$                      & $9.8\times 10^3$   & $2.0\times 10^5$     \\
   2f                   &   $1.2\times 10^5$    & $2.4\times 10^8$   & $2.3\times 10^6$ 
                         &   $1.4\times 10^4$    & $2.9\times 10^7$   & $2.3\times 10^6$     \\   
   4f                   &   $2.7\times 10^4$    & $5.4\times 10^7$   & $6.9\times 10^6$ 
                         &   $1.3\times 10^4$    & $2.6\times 10^7$   & $6.9\times 10^6$    \\
   6f                   &   $6.9\times 10^2$    & $1.4\times 10^6$   & $5.0\times 10^6$ 
                         &   $2.4\times 10^2$    & $4.8\times 10^5$   & $5.0\times 10^6$    \\
   \hline
   \end{tabular}
  \caption{The cross sections ($\sigma$), the number of expected events ($N_{\rm{exp}}$), and the number of generated events ($N_{\rm{gen}}$) for signal and background processes for left-handed (eLpR) and 
  right-handed (eRpL) beam polarisations as defined in the text.}
\label{tab:xsec}
  \end{center}
\end{table}

\subsection{Event selection and results}
\subsubsection{Pre-selection}
The general characteristics of signal events are an isolated hard photon with energy $E_\gamma\sim 220$ GeV (namely primary photon), and a large recoil mass ($M_{rec}$) against that primary photon with $M_{rec}\sim M_X$. In the pre-selection, among all the photons identified by PandorPFA, the photon with energy closest to 220 GeV is selected as the primary photon candidate, and its energy is required to be larger than 50 GeV. Due to the fact that a hard photon is possibly reconstructed as several separated clusters by PandoraPFA, a merging procedure is carried out so that any photons within a small cone (cosine of the cone angle is 0.998) around the selected primary photon candidate are merged into the candidate with total four momentum equals to the sum of four momenta of all photons in that cone. To suppress background events with such a photon candidate from jets, the selected candidate photon is required to satisfy the isolation criterion, $\frac{E_{\rm{cone}}}{E_\gamma}<5\%$, where $E_{\rm{cone}}$ is the sum of energies from charged particles in a relatively large cone around the primary photon with cosine of the cone angle $\cos\theta_{\rm{cone}}=0.98$, and is then selected as the primary photon.

\subsubsection{Inclusive mode}
\begin{figure}[ht]
  \centering
  \begin{tabular}[c]{cc}
    \includegraphics[width=0.5\columnwidth]{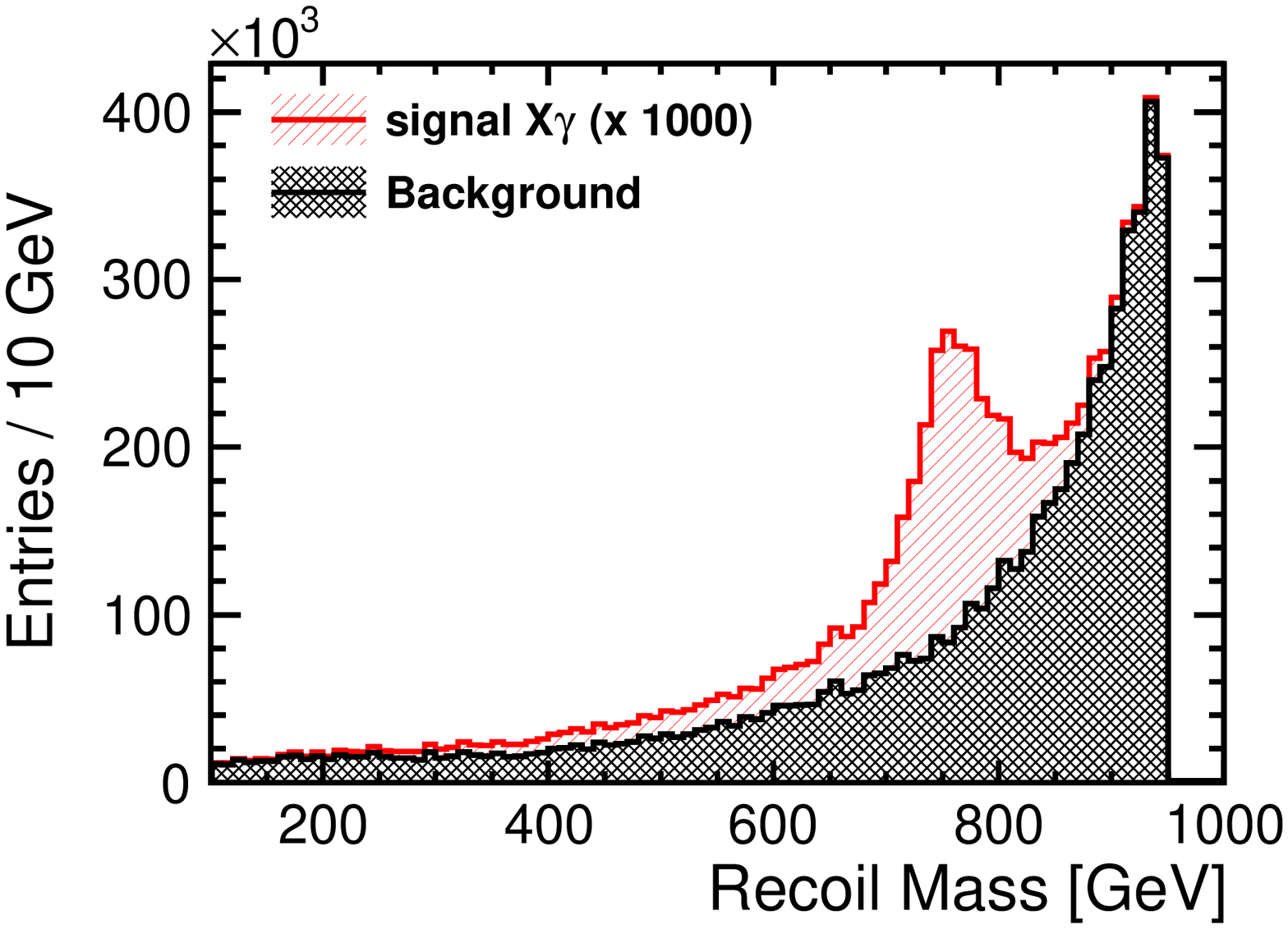} &
    \includegraphics[width=0.5\columnwidth]{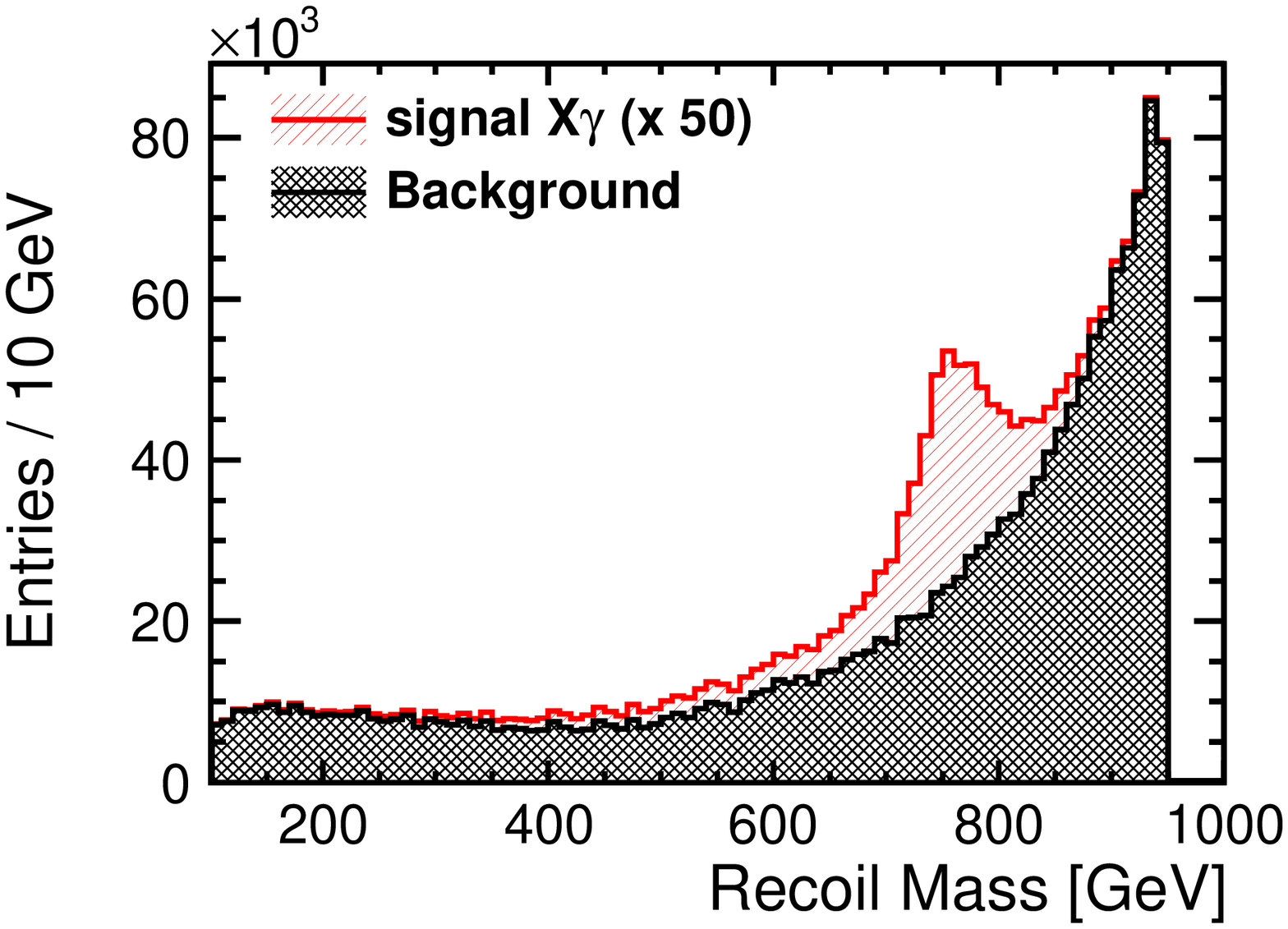} \\   
 \end{tabular}
  \caption{In inclusive search mode, the stacked distributions of recoil mass against $\gamma$ after all cuts for signal events (hatched, in red) and total background events (tiled, in black) in the cases of left-handed beam polarisations (left, signal is scaled by a factor of 1000) and right-handed beam polarisations (right, signal is scaled by a factor of 50).}
  \label{fig:mrecoil}
\end{figure}

In the inclusive mode, most of the background events that survive after the pre-selection are those with one hard photon from ISR. Since the ISR photons are mostly in the forward or backward direction with respects to the electron beam direction, to further suppress the background events, one additional selection cut is applied, which is $|\cos\theta_\gamma|<0.88$, where $\theta_\gamma$ is the polar angle of the selected primary photon. The spectrum of recoil mass against the primary photon is shown in Fig.~\ref{fig:mrecoil} as a stacked histogram for remaining signal (in red) and all background (in black) events. For the purpose to visualise the signal shape, the signal component is scaled by a factor of 10 or larger depending on the decay modes throughout all the plots here and after. The number of signal and background events before and after each selection cut together with significances are shown in Table~\ref{tab:reduction_recoil}. The efficiencies for background processes are all similar because the ISR effect is not much process dependent. The significance ($n_{\rm{sig}}$) after final selection is calculated as $n_{\rm{sig}}^2=\sum_i\frac{S_i^2}{S_i+B_i}$, where $S_i$ and $B_i$ are the number of signal and background events respectively in bin $i$ of the recoil mass spectrum, and the summation of $i$ goes over all the bins from 300 GeV to 900 GeV. In the inclusive mode, the final significances for the left and right handed beam polarisations are $1.6\sigma$ and $9.6\sigma$, respectively. 

\begin{table}[ht]
 \begin{center}
 \begin{tabular}{| l | r | r | r | r | r | r |}
  \hline
  \multirow{2}{*}{Process}          &   \multicolumn{3}{c|}{eLpR}          &   \multicolumn{3}{c|}{eRpL}  \\
          \cline{2-7}  
                          &  before selection    & pre-selection           & $|\cos\theta_\gamma|<0.88$                
                          &  before selection    & pre-selection           & $|\cos\theta_\gamma|<0.88$      \\                                    
   \hline
   signal             &  $3.1\times 10^3$   & $2.9\times 10^3$    & $2.4\times 10^3$    
                         &  $9.8\times 10^3$   & $9.1\times 10^3$    & $7.6\times 10^3$    \\   
   background   &  $3.0\times 10^8$    & $1.6\times 10^7$    & $6.0\times 10^6$     
                         &  $5.6\times 10^7$    & $5.2\times 10^6$    & $1.6\times 10^6$   \\    
   \hline           
   2f                   &  $2.4\times 10^8$    & $1.1\times 10^7$    & $4.8\times 10^6$      
                         &  $2.9\times 10^7$    & $2.9\times 10^6$    & $1.1\times 10^6$    \\      
   4f                   &  $5.4\times 10^7$    & $4.7\times 10^6$    & $1.1\times 10^6$          
                         &  $2.6\times 10^7$    & $2.3\times 10^6$    & $5.7\times 10^5$    \\         
   6f                   &  $1.4\times 10^6$    & $9.7\times 10^4$    & $3.4\times 10^4$             
                         &  $4.8\times 10^5$    & $3.6\times 10^4$    & $1.3\times 10^4$    \\            
   \hline
   significance    &   \multicolumn{3}{c|}{$1.6\sigma$}                
                          &   \multicolumn{3}{c|}{$9.6\sigma$}    \\               
   \hline
   \end{tabular}
  \caption{For inclusive mode, the number of signal and background events after each selection cut together with final significances. The selection cuts are explained in the text. Results for both left and right handed beam polarisations are shown.}
\label{tab:reduction_recoil}
  \end{center}
\end{table}

\subsubsection{$X\to\rm{invisible}$ mode}
\begin{figure}[ht]
  \centering
  \begin{tabular}[c]{cc}
    \includegraphics[width=0.5\columnwidth]{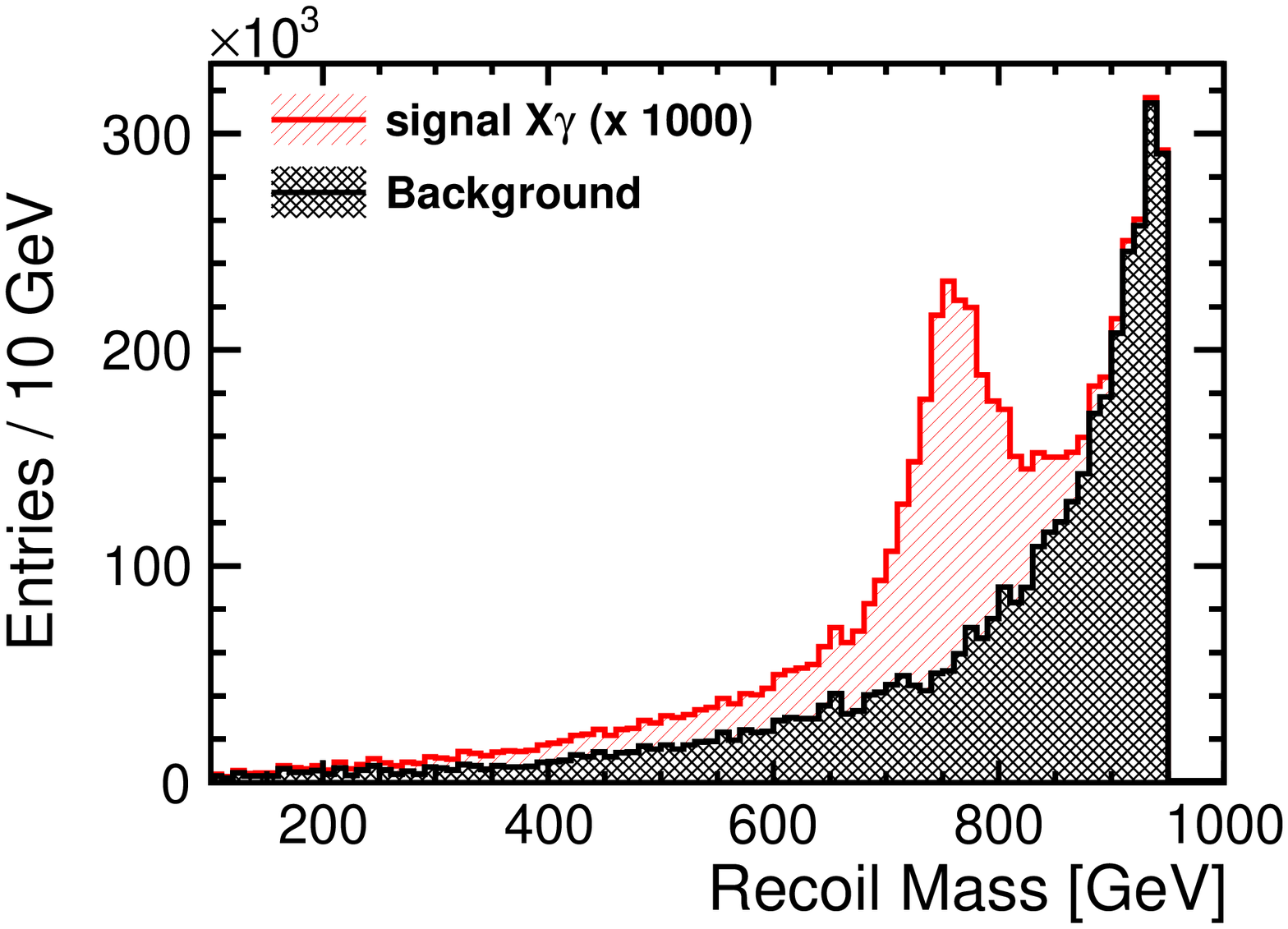} &
    \includegraphics[width=0.5\columnwidth]{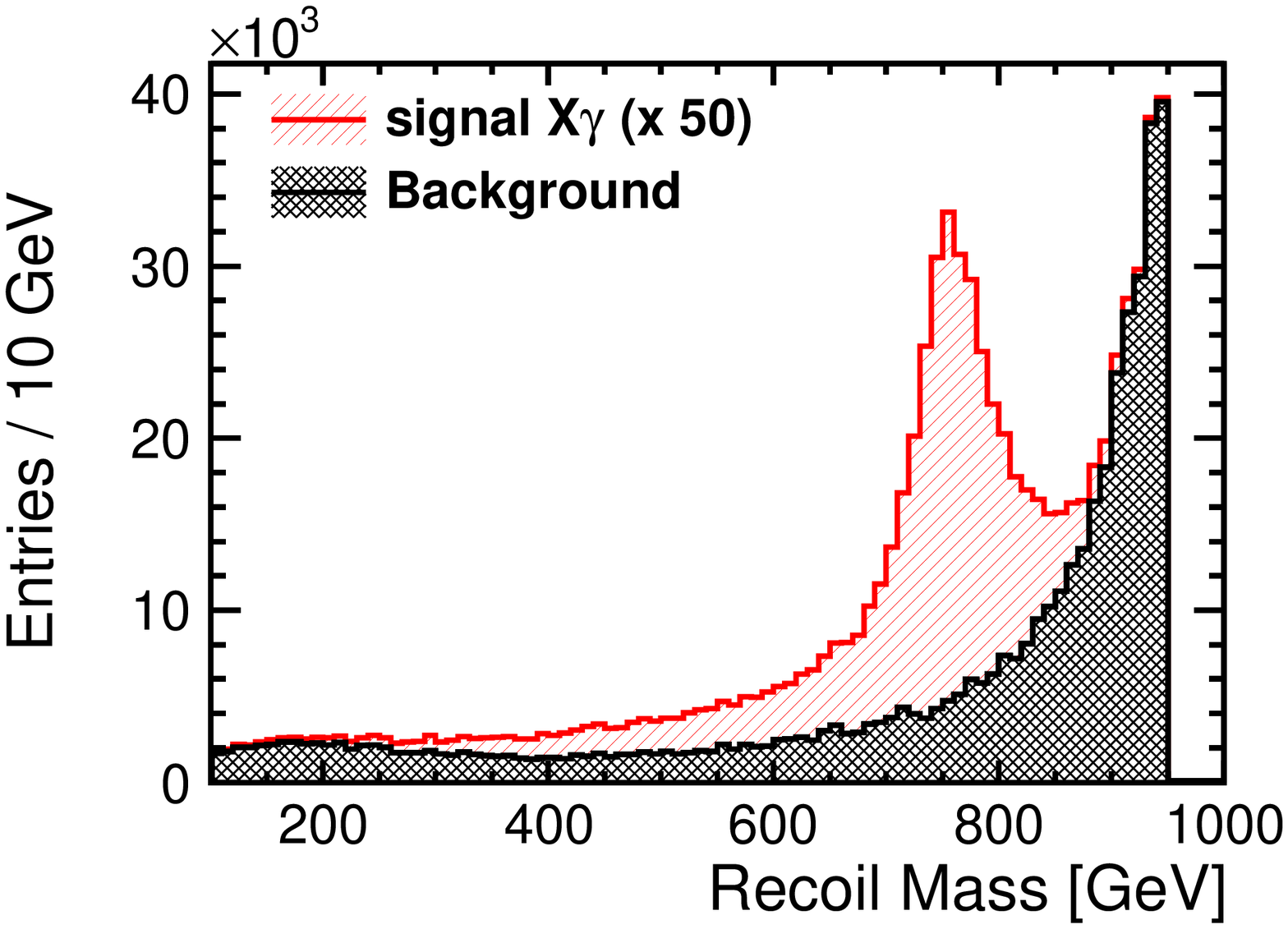} \\   
 \end{tabular}
  \caption{In the case of $X\to\mathrm{invisible}$ decays, the stacked distributions of recoil mass against $\gamma$ after all cuts for signal events (hatched, in red) and total background events (tiled, in black) in the cases of eLpR (left, signal is scaled by a factor of 1000) and eRpL (right, signal is scaled by a factor of 50).}
  \label{fig:mrecoil_inv}
\end{figure}

In the $X\to\rm{invisible}$ mode, first of all the same cut $|\cos\theta_\gamma|<0.88$ is applied after the pre-selection as in the inclusive mode to suppress the background events with a forward or backward hard ISR photon. In addition, since there is no expected visible particle from $X$ reconstructed in the detector, a cut on the visible energy ($E_{\rm{vis}}$) is applied, $E_{\rm{vis}}<60~\rm{GeV}$, where $E_{\rm{vis}}$ is the sum of energies from all charged final state particles. Here the usage of charge particles only is because there are possibly additional ISR photons in each signal event, and the relatively large cut value of 60 GeV is because of the pile-up of beam induced background events. The spectrum of recoil mass against the primary photon after all selection cuts is shown in Fig.~\ref{fig:mrecoil_inv}. The background is dominated by $e^+e^-\to\nu_e\bar{\nu}_e\gamma$ for the left-handed beam polarisations, and for the right handed beam polarisations is dominated by $e^+e^-\to\nu\bar{\nu}\gamma$ for all three flavours of neutrinos. The number of signal and background events remaining after each selection cut are shown in Table~\ref{tab:reduction_inv}. In the $X\to\rm{invisible}$ mode, the final significances for the left and right handed beam polarisations are $2.0\sigma$ and $20\sigma$, respectively.

\begin{table}[ht]
\scriptsize
 \begin{center}
 \begin{tabular}{| l | r | r | r | r | r | r | r | r |}
  \hline
  \multirow{2}{*}{Process}          &   \multicolumn{4}{c|}{eLpR}          &   \multicolumn{4}{c|}{eRpL}  \\
          \cline{2-9}  
               &  before selection    & pre-selection    & $|\cos\theta_\gamma|<0.88$  & $E_{\rm{vis}}<60~\rm{GeV}$                
               &  before selection    & pre-selection    & $|\cos\theta_\gamma|<0.88$  & $E_{\rm{vis}}<60~\rm{GeV}$    \\                                    
   \hline
   signal             &  $3.1\times 10^3$   & $2.9\times 10^3$    & $2.4\times 10^3$   & $2.3\times 10^3$ 
                         &  $9.8\times 10^3$   & $9.1\times 10^3$    & $7.6\times 10^3$   & $7.4\times 10^3$  \\   
   background   &  $3.0\times 10^8$    & $1.6\times 10^7$    & $6.0\times 10^6$  & $4.0\times 10^6$     
                         &  $5.6\times 10^7$    & $5.2\times 10^6$    & $1.6\times 10^6$  & $4.6\times 10^5$    \\    
   \hline           
   2f                   &  $2.4\times 10^8$    & $1.1\times 10^7$    & $4.8\times 10^6$  & $3.8\times 10^6$      
                         &  $2.9\times 10^7$    & $2.9\times 10^6$    & $1.0\times 10^6$  & $3.8\times 10^5$    \\      
   4f                   &  $5.4\times 10^7$    & $4.7\times 10^6$    & $1.1\times 10^6$  & $1.0\times 10^5$          
                         &  $2.6\times 10^7$    & $2.3\times 10^6$    & $5.7\times 10^5$  & $7.3\times 10^4$    \\         
   6f                   &  $1.4\times 10^6$    & $9.7\times 10^4$    & $3.4\times 10^4$  & $1.6\times 10^2$             
                         &  $4.8\times 10^5$    & $3.6\times 10^4$    & $1.3\times 10^4$  & $4.3\times 10^1$    \\            
   \hline
   significance    &   \multicolumn{4}{c|}{$2.0\sigma$}                
                          &   \multicolumn{4}{c|}{$20\sigma$}    \\               
   \hline
   \end{tabular}
  \caption{For $X\to\rm{invisible}$ mode, the number of signal and background events after each selection cut together with final significances. The selection cuts are explained in the text. Results for both left and right handed beam polarisations are shown.}
\label{tab:reduction_inv}
  \end{center}
\end{table}

\subsubsection{$X\to b\bar{b}$ mode}
\begin{figure}[ht]
  \centering
  \begin{tabular}[c]{cc}
    \includegraphics[width=0.5\columnwidth]{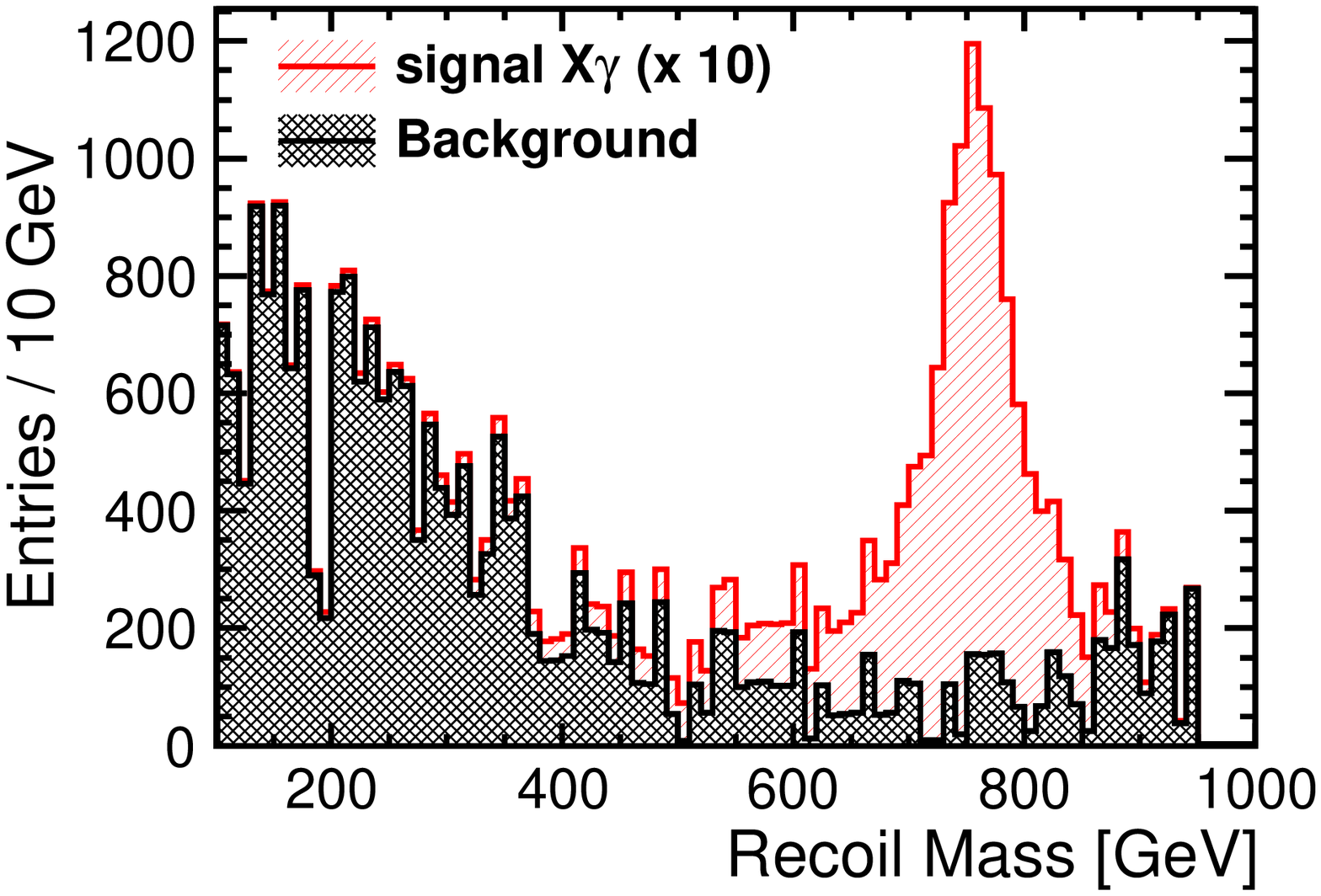} &
    \includegraphics[width=0.5\columnwidth]{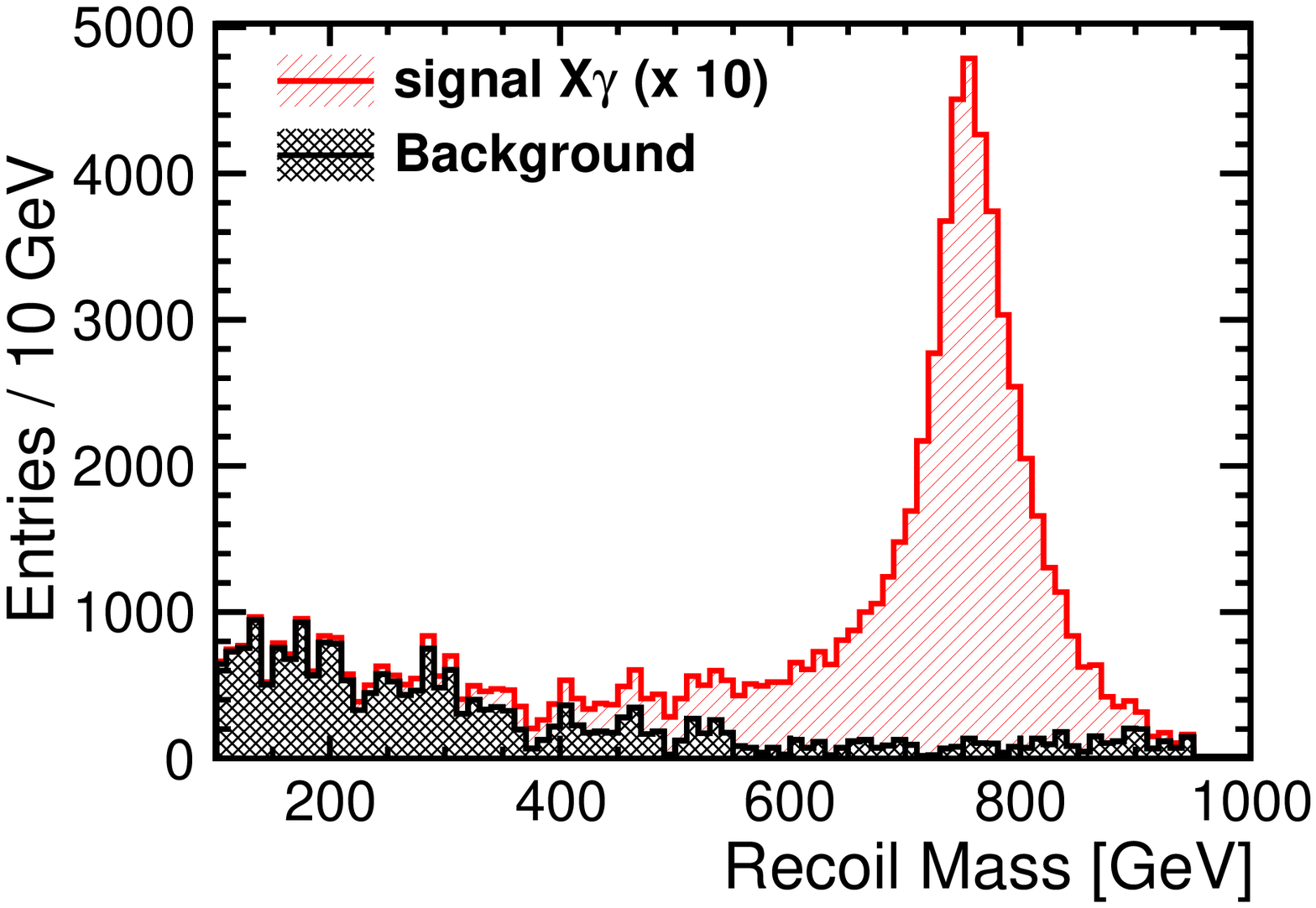} \\   
 \end{tabular}
  \caption{For $X\to b\bar{b}$, the stacked distributions of recoil mass against $\gamma$ after all cuts for signal events (hatched, in red) and total background events (tiled, in black) in the cases of eLpR (left, signal is scaled by a factor of 10) and eRpL (right, signal is scaled by a factor of 10).}
  \label{fig:mrecoil_bb}
\end{figure}

In the $X\to b\bar{b}$ mode, all the reconstructed particles, except those already selected as the primary photon, are used as input to 2-jets clustering using longitudinal invariant $k_T$ algorithm~\cite{KT_JET} implemeted in the FastJet package~\cite{FastJet} with $R=1.5$. This step is effective to remove some of the particles which come from the pile-up. The particles remaining after this step are clustered into two jets using Durham jet algorithm~\cite{Durham}. Each of the two jets is then flavour tagged using LCFIPlus. The output of flavour tagging by LCFIPlus for each jet is a value which is a measure of likelihood that that jet is a $b$-jet, and here the output values for the two jets are called $btag_1$ and $btag_2$, with $btag_1>btag_2$. The following selection cuts are then applied to every event in addition to the pre-selection for the $X\to b\bar{b}$ mode:
\begin{enumerate}[i.)]
 \item $N_{c}>5$ for each jet, where $N_c$ is the number of charged particles in the jet. This cut is effective to suppress background events with fewer number of particles in the final state such as lepton pair events.
 \item $Y_{4\to 3}<0.002~(0.002)$ and $E_{\rm{mis}}<260~(180)~\rm{GeV}$, where $Y_{4\to 3}$ is the smallest $Y_{ij}$ used in Durham jet clustering from 4-jet to 3-jet, and $E_{\rm{mis}}$ is the missing energy. Hereafter the values are optimised corresponding to right-handed (left-handed) beam polarisations. The $Y_{4\to 3}$ cut is effective to suppress background events with more than 3 primary partons such as full hadronic modes of $4\rm{f}$ and $6\rm{f}$ events.
 \item $btag_1>0.74~(0.77)$ and $btag_2>0.1~(0.3)$, which are effective to suppress background events without $b$-jets such as light quark pair events.
 \item $|\cos\theta_\gamma|<0.95~(0.88)$, which is effective to suppress all background events with small angle ISR photons.
 \item $M_{bb}>350~(530)~\rm{GeV}$, where $M_{bb}$ is the invariant mass of the two jets. Since $M_{bb}$ is highly correlated to $M_{rec}$, this cut just provides small improvement to the overall background suppression.
\end{enumerate} 
The spectrum of recoil mass against the primary photon after all the selection cuts except the last cut on $M_{bb}$ is shown in Fig.~\ref{fig:mrecoil_bb}. The background is dominated by $e^+e^-\to b\bar{b}\gamma$. The number of signal and background events remaining after each selection cut is shown in Table~\ref{tab:reduction_bb_l} for the left-handed beam polarisations and in Table~\ref{tab:reduction_bb_r} for the right-handed beam polarisations. In the $X\to b\bar{b}$ mode, the final significances for the left and right handed beam polarisations are $23\sigma$ and $62\sigma$, respectively.

\begin{table}[ht]
\footnotesize
 \begin{center}
 \begin{tabular}{| l | r | r | r | r | r | r | r | }
  \hline
   \multirow{2}{*}{Process}       &  \multirow{2}{*}{before selection}  & \multirow{2}{*}{pre-selection} 
& \multirow{2}{*}{$N_j>5$}    & $Y_{4\to 3}<0.002$ & $btag_1>0.77$ 
& \multirow{2}{*}{$|\cos\theta_\gamma|<0.88$}  & \multirow{2}{*}{$M_{bb}>530~\rm{GeV}$} \\
  &  &  &  & $E_{\rm{mis}}<180~\rm{GeV}$ & $btag_2>0.3$ & &  \\
   \hline
   signal             &  $3.1\times 10^3$   & $2.7\times 10^3$    & $2.6\times 10^3$   & $2.0\times 10^3$    
                         &  $1.5\times 10^3$   & $1.3\times 10^3$    & $1.1\times 10^3$  \\   
   background   &  $3.0\times 10^8$    & $1.3\times 10^7$    & $1.2\times 10^6$  & $4.4\times 10^5$     
                         &  $5.2\times 10^4$    & $2.8\times 10^4$    & $3.5\times 10^3$  \\    
   \hline           
   2f                   &  $2.4\times 10^8$    & $8.1\times 10^6$    & $5.0\times 10^5$  & $3.1\times 10^5$      
                         &  $5.0\times 10^4$    & $2.6\times 10^4$    & $2.6\times 10^3$   \\      
   4f                   &  $5.4\times 10^7$    & $4.6\times 10^6$    & $6.0\times 10^5$  & $1.3\times 10^5$          
                         &  $1.6\times 10^3$    & $6.4\times 10^2$    & $3.2\times 10^2$   \\         
   6f                   &  $1.4\times 10^6$    & $9.7\times 10^4$    & $5.2\times 10^4$  & $4.8\times 10^3$             
                         &  $1.3\times 10^3$    & $7.0\times 10^2$    & $5.6\times 10^2$   \\            
   \hline
   significance    &   \multicolumn{7}{c|}{$23\sigma$}    \\            
   \hline
   \end{tabular}
  \caption{For $X\to b\bar{b}$ mode and eLpR, the number of signal and background events after each selection cut together with final significances. The selection cuts are explained in the text.}
\label{tab:reduction_bb_l}
  \end{center}
\end{table}

\begin{table}[ht]
\footnotesize
 \begin{center}
 \begin{tabular}{| l | r | r | r | r | r | r | r | }
  \hline
   \multirow{2}{*}{Process}       &  \multirow{2}{*}{before selection}  & \multirow{2}{*}{pre-selection} 
& \multirow{2}{*}{$N_j>5$}    & $Y_{4\to 3}<0.002$ & $btag_1>0.74$ 
& \multirow{2}{*}{$|\cos\theta_\gamma|<0.95$}  & \multirow{2}{*}{$M_{bb}>350~\rm{GeV}$} \\
  &  &  &  & $E_{\rm{mis}}<260~\rm{GeV}$ & $btag_2>0.1$ & &  \\
   \hline
   signal             &  $9.8\times 10^3$   & $8.6\times 10^3$    & $8.2\times 10^3$   & $7.2\times 10^3$    
                         &  $6.1\times 10^3$   & $5.7\times 10^3$    & $5.4\times 10^3$  \\   
   background   &  $5.6\times 10^7$    & $4.8\times 10^6$    & $4.0\times 10^5$  & $2.3\times 10^5$     
                         &  $4.3\times 10^4$    & $2.8\times 10^4$    & $4.1\times 10^3$  \\    
   \hline           
   2f                   &  $2.9\times 10^7$    & $2.6\times 10^6$    & $2.9\times 10^5$  & $2.1\times 10^5$      
                         &  $4.0\times 10^4$    & $2.7\times 10^4$    & $2.9\times 10^3$   \\      
   4f                   &  $2.6\times 10^7$    & $2.2\times 10^6$    & $9.6\times 10^4$  & $2.3\times 10^4$          
                         &  $1.8\times 10^3$    & $6.5\times 10^2$    & $4.5\times 10^2$   \\         
   6f                   &  $4.8\times 10^5$    & $3.6\times 10^4$    & $2.0\times 10^4$  & $2.6\times 10^3$             
                         &  $1.1\times 10^3$    & $7.4\times 10^2$    & $6.7\times 10^2$   \\            
   \hline
   significance    &   \multicolumn{7}{c|}{$62\sigma$}    \\            
   \hline
   \end{tabular}
  \caption{For $X\to b\bar{b}$ mode and eRpL, the number of signal and background events after each selection cut together with final significances. The selection cuts are explained in the text.}
\label{tab:reduction_bb_r}
  \end{center}
\end{table}

\section{Discussion and Conclusion} 
\label{Sec:summary}

We have performed a realistic simulation for $e^+e^-\to X\gamma$ with
full detector simulation. 
We have shown that assuming an integrated luminosity of
2~$\rm{ab}^{-1}$ at the ILC, without assuming the decay modes, the cross
section of $e^+e^-\to X\gamma$ for eRpL
($\sigma(X\gamma)_R$) can be measured to 10\%. 
In the case that the decay is dominated by $X\to\rm{invisible}$, using
the right-handed beam polarisations
$\sigma(X\gamma)_R\times{\rm{BR}}(X\to\rm{invisible})$ can be measured
to 5\%.
If $X\to b\bar{b}$ dominates the decay,
$\sigma(X\gamma)_R\times{\rm{BR}}(X\to b\bar{b})$ can be measured to
1.6\% and $\sigma(X\gamma)_L\times{\rm{BR}}(X\to b\bar{b})$ (for eLpR)
 can be measured to 4.3\%.
${\rm{BR}}(X\to b\bar{b})$ can be extracted by two measurements of
$\sigma(X\gamma)_R$ and $\sigma(X\gamma)_R\times{\rm{BR}}(X\to
b\bar{b})$, which together with $\sigma(X\gamma)_L\times{\rm{BR}}(X\to
b\bar{b})$ can be used to further extract $\sigma(X\gamma)_L$.
Then the absolute values of the effective couplings to ${\rm{SU(2)}}_L$ and ${\rm{U(1)}}_Y$
fields can be measured separately. 

The results can be also given in terms of the range of
 $\Gamma_{\gamma\gamma}$ that allows ILC to detect $X$ with more than 
$5\sigma$ significance.
Interestingly enough, if $X\to b\bar{b}$ is the dominant decay mode, the
range $\Gamma_{\gamma\gamma}>30$~MeV
($\Gamma_{\gamma\gamma}/M_X>4\times10^{-5}$) can be explored at the ILC,
which covers the full allowed range if the diphoton resonance is
dominantly produced by $b\bar{b}$ fusion at the
LHC~\cite{Franceschini:2015kwy}.

To summarise, we have investigated the prospects of diphoton resonance
studies in $e^+e^-\to X\gamma/XZ$ at the ILC.
Within the framework of the effective Lagrangian for spin-0 hypothesis
we have investigated the production cross sections as well as the
angular distributions of the processes, and found that these are useful to
determine the properties of the resonance, such as its mass, spin,
parity, total decay-width, and absolute values of effective couplings to
the SU(2)$_L$ and U(1)$_Y$ gauge fields.
A realistic simulation for $e^+e^-\to X\gamma$ assuming the large decay
width has been performed based on full detector simulation. 
Complementary to the searches at the LHC, the absolute values of 
production cross-sections can be measured at the ILC using
recoil technique without assuming decay modes of $X$.
In addition, we have shown that the measurement at the ILC is
capable of the searches for $X\to\rm{invisible}$ and $X\to b\bar{b}$
decays with high sensitivities. It should be emphasised that the studies 
presented here are generic and applicable to any similar new particles
which couple to $\gamma\gamma$ even if the diphoton resonance turns
out to be a statistical fluctuation, and hence remain useful.

\section*{Note Added}

During this manuscript was under review, new experimental data which
were collected at the LHC in 2016 were analyzed by the ATLAS and CMS
Collaborations, and the diphoton resonance at $M_X\sim750$~GeV suggested
by the 2015 data turned out to be most likely a statistical
flucturation~\cite{ATLAS:2016eeo,CMS:2016crm}. 
Although numerical studies done in this paper focus on the 750~GeV
resonance as a case study, as we have stated in the main text, our
analysis is applicable to any resonance which couples to diphoton.
If there still exists a resonance with relatively low mass~(several
hundreds GeV), the resonance should have small branching ratio to
diphoton or other dibosons in order to avoid the LHC constraints.
On the other hand, to have sizable cross section at the ILC, the
resonance should have large coupling to dibosons. 
Therefore, the resonance which could be explored at the ILC has a large
decay width with small branching ratios to dibosons, but the other modes
such as the invisible decay or the $b\bar b$ decay modes may dominate.
Thus, the search strategy which we have studied in this paper is
useful for the future possible resonance by adjusting the detailed
numbers of signal and background rates accordingly.
In the case where a resoance is heavier than the collision energy,
the cross section is expected to be small due to the off-shell
suppression. 
We refer the readers to the comprehensive analysis of the indirect
searches for the physics related to the diphoton
resonance~\cite{Fujii:2016raq}.

\acknowledgments
We are grateful to Michael Peskin and Francois Richard for valuable
discussions.
H.Y.\ thanks to Pyungwon Ko, Kentarou Mawatari, Kenji Nishiwaki and
Takaaki Nomura for useful discussions. 
K.F.\ and J.T.\ thank to the ILD Concept Group for providing common
simulation and analysis framework used in this paper; in particular,
T.~Barklow, A.~Miyamoto, and M.~Berggren for their work in generating
the background Monte-Carlo samples, and J.~List for useful discussions
for the analysis. J.T.\ thanks to C.~Calancha for developing the
database tool to effectively get cross section values for background
processes.

\appendix

\section{Partial decay widths of $X$}
In this Appendix, we present the partial decay widths of $X$ into pairs
of SM gauge bosons based on the effective Lagrangians in
Eq.~(\ref{eq:lag0}) and (\ref{eq:lagP}).
For the scalar case, the partial decay widths are calculated to
be
\begin{align}
 & \Gamma_{\gamma\gamma} =
 \frac{M_X^3}{64\pi}\left(\frac{c_\gamma}{\Lambda}\right)^2,\\
 & \Gamma_{\gamma Z} = \frac{M_X^3}{128\pi}\left(\frac{c_{\gamma
 Z}}{\Lambda}\right)^2\left(1-\frac{M_Z^2}{M_X^2}\right)^3, \\
 & \Gamma_{ZZ} =
 \frac{M_X^3}{64\pi}\left(\frac{c_Z}{\Lambda}\right)^2
 \beta_Z\left(1-4\frac{M_Z^2}{M_X^2}+6\frac{M_Z^4}{M_X^4}\right), \\
 & \Gamma_{WW} =
 \frac{M_X^3}{128\pi}\left(\frac{c_W}{\Lambda}\right)^2
 \beta_W\left(1-4\frac{M_W^2}{M_X^2}+6\frac{M_W^4}{M_X^4}\right), \\
 & \Gamma_{gg} =
 \frac{M_X^3}{8\pi}\left(\frac{c_g}{\Lambda}\right)^2,
\end{align}
where $\beta_V=\sqrt{1-4M_V^2/M_X^2}$ for $V=W$, $Z$.

For the pseudoscalar case, the partial decay widths are calculated to be
\begin{align}
 & \Gamma_{\gamma\gamma} =
 \frac{M_X^3}{64\pi} \left(\frac{\tilde c_\gamma}{\Lambda}\right)^2, \\
 & \Gamma_{\gamma Z} =
 \frac{M_X^3}{128\pi}\left(\frac{\tilde c_{\gamma
 Z}}{\Lambda}\right)^2 \left(1-\frac{M_Z^2}{M_X^2}\right)^3, \\
 & \Gamma_{ZZ} =
 \frac{M_X^3}{64\pi}\left(\frac{\tilde c_Z}{\Lambda}\right)^2\beta_Z^3, \\
 & \Gamma_{WW} =
 \frac{M_X^3}{128\pi}\left(\frac{\tilde c_W}{\Lambda}\right)^2\beta_W^3, \\
 & \Gamma_{gg} =
 \frac{M_X^3}{8\pi} \left(\frac{\tilde c_g}{\Lambda}\right)^2.
\end{align}

\section{Lepton angular distributions}
The lepton angular distribution in the decay of $Z\to\ell^-\ell^+$ in
$e^+e^-\to XZ$ is expressed by using the 6 structure functions
$F_{i}(s,\cos\theta)$ for $i=1$ to 6 as
\begin{align}
{\mathcal D}(\lambda_V;\theta,\hat\theta,\hat\phi) = 
 \frac{9}{128\pi{\mathcal F}} & \left[ F_1(1 + \cos^2\hat\theta) +
 F_2(1-3\cos^2\hat\theta) + F_3\sin2\hat\theta\cos\hat\phi
 + F_4\sin^2\hat\theta\cos2\hat\phi
\right. \nonumber \\ & \left.
 + F_5\cos\hat\theta + F_6\sin\hat\theta\cos\hat\phi \right].
\end{align}

For the scalar case,
\begin{align}
 & F_1 = \left(\beta_Z^2+\frac{4M_Z^2}{s}\right)(1+\cos^2\theta) +
 \frac{4M_Z^2}{s}\sin^2\theta, 
 \quad F_2 = \frac{4M_Z^2}{s}\sin^2\theta, \nonumber \\
 & F_3 = -\frac{4M_Z}{\sqrt{s}}
 \sqrt{\beta_Z^2+\frac{4M_Z^2}{s}}\cos\theta\sin\theta,
 \quad
 F_4 = \left(\beta_Z^2+\frac{4M_Z^2}{s}\right)\sin^2\theta,
 \nonumber \\ & F_5 =
 4\lambda_V\xi\left(\beta_Z^2+\frac{4M_Z^2}{s}\right)\cos\theta,
 \quad F_6 =
 -8\lambda_V\xi\frac{M_Z}{\sqrt{s}}
 \sqrt{\beta_Z^2+\frac{4M_Z^2}{s}}\sin\theta,
\end{align}
with ${\mathcal F}=\beta_Z^2+6M_Z^2/s$ and $\xi$ is defined in
Eq.~(\ref{eq:xi}).

For the pseudoscalar case, ${\mathcal F}=1$ and 
\begin{align}
 F_1 = 1+\cos^2\theta, \quad F_2 = F_3 = F_6 = 0, \quad F_4 =
 -\sin^2\theta, \quad F_5 = 4\lambda_V\xi\cos\theta. 
\end{align}
%



\begin{thebibliography}{99}

\bibitem{Aaboud:2016tru} 
  M.~Aaboud {\it et al.} [ATLAS Collaboration],
  arXiv:1606.03833 [hep-ex].

\bibitem{Khachatryan:2016hje} 
  V.~Khachatryan {\it et al.} [CMS Collaboration],
  arXiv:1606.04093 [hep-ex].


\bibitem{Aad:2015mna} 
  G.~Aad {\it et al.} [ATLAS Collaboration],
  Phys.\ Rev.\ D {\bf 92}, no. 3, 032004 (2015).

\bibitem{Khachatryan:2015qba} 
  V.~Khachatryan {\it et al.} [CMS Collaboration],
  Phys.\ Lett.\ B {\bf 750}, 494 (2015).

\bibitem{Franceschini:2015kwy} 
  R.~Franceschini {\it et al.},
  JHEP {\bf 1603}, 144 (2016).
  
\bibitem{Strumia:2016wys} 
  A.~Strumia,
  arXiv:1605.09401 [hep-ph].
	
\bibitem{Franceschini:2016gxv} 
  R.~Franceschini, G.~F.~Giudice, J.~F.~Kamenik, M.~McCullough, F.~Riva, A.~Strumia and R.~Torre,
  arXiv:1604.06446 [hep-ph].
	

\bibitem{Baer:2013cma} 
  H.~Baer {\it et al.},
  arXiv:1306.6352 [hep-ph].

\bibitem{Fujii:2015jha} 
  K.~Fujii {\it et al.},
  arXiv:1506.05992 [hep-ex].

\bibitem{Djouadi:2016eyy} 
  A.~Djouadi, J.~Ellis, R.~Godbole and J.~Quevillon,
  JHEP {\bf 1603}, 205 (2016).

\bibitem{Richard:2016nhm} 
  F.~Richard,
  arXiv:1604.01640 [hep-ex].

\bibitem{Ito:2016kvw} 
  H.~Ito and T.~Moroi,
  arXiv:1604.04076 [hep-ph].

\bibitem{Bae:2016oey} 
  K.~J.~Bae, K.~Hamaguchi, T.~Moroi and K.~Yanagi,
  Phys.\ Lett.\ B {\bf 759}, 575 (2016).

\bibitem{Ito:2016zkz} 
  H.~Ito, T.~Moroi and Y.~Takaesu,
  Phys.\ Lett.\ B {\bf 756}, 147 (2016).
  
\bibitem{He:2016olo} 
  M.~He, X.~G.~He and Y.~Tang,
  Phys.\ Lett.\ B {\bf 759}, 166 (2016).
  	
\bibitem{Giddings:2016sfr} 
  S.~B.~Giddings and H.~Zhang,
  Phys.\ Rev.\ D {\bf 93}, no. 11, 115002 (2016).


\bibitem{Arun:2015ubr} 
  M.~T.~Arun and P.~Saha,
  arXiv:1512.06335 [hep-ph].

\bibitem{Han:2015cty} 
  C.~Han, H.~M.~Lee, M.~Park and V.~Sanz,
  Phys.\ Lett.\ B {\bf 755}, 371 (2016).

\bibitem{Martini:2016ahj} 
  A.~Martini, K.~Mawatari and D.~Sengupta,
  Phys.\ Rev.\ D {\bf 93}, no. 7, 075011 (2016).
  
\bibitem{Geng:2016xin} 
  C.~Q.~Geng and D.~Huang,
  Phys.\ Rev.\ D {\bf 93}, no. 11, 115032 (2016).

\bibitem{Sanz:2016auj} 
  V.~Sanz,
arXiv:1603.05574 [hep-ph].

\bibitem{Bernon:2016dow} 
  J.~Bernon, A.~Goudelis, S.~Kraml, K.~Mawatari and D.~Sengupta,
  JHEP {\bf 1605}, 128 (2016).

\bibitem{CMS:2015neg} 
  CMS Collaboration [CMS Collaboration],
  CMS-PAS-EXO-14-005.

\bibitem{Alves:2015jgx} 
  A.~Alves, A.~G.~Dias and K.~Sinha,
  Phys.\ Lett.\ B {\bf 757}, 39 (2016).

\bibitem{Physsim}
K.~Fujii, et al., http://www-jlc.kek.jp/subg/offl/physsim/

\bibitem{Ko:2016xwd} 
  P.~Ko and H.~Yokoya,
  arXiv:1603.04737 [hep-ph].

\bibitem{Whizard}
W. Kilian, T. Ohl, and J. Reuter, Eur. Phys. J. C71, 1742 (2011).

\bibitem{Pythia} 
  T.~Sjostrand, S.~Mrenna and P.~Z.~Skands,
  JHEP {\bf 0605}, 026 (2006).
  
\bibitem{GuineaPig} 
  D.~Schulte,
  DESY-TESLA-97-08, TESLA-97-08.

\bibitem{ILCTDR} 
  C.~Adolphsen {\it et al.},
  arXiv:1306.6328 [physics.acc-ph].

\bibitem{LLA} 
  M.~Skrzypek and S.~Jadach,
  Z.\ Phys.\ C {\bf 49}, 577 (1991).

\bibitem{TDRVol4} 
  T.~Behnke {\it et al.},
  arXiv:1306.6329 [physics.ins-det].

\bibitem{GEANT4} 
  S.~Agostinelli {\it et al.} [GEANT4 Collaboration],
  Nucl.\ Instrum.\ Meth.\ A {\bf 506}, 250 (2003).

 \bibitem{ILD} 
  T.~Abe {\it et al.} [Linear Collider ILD Concept Group - Collaboration],
  arXiv:1006.3396 [hep-ex].

\bibitem{Mokka}
 P. Mora de Freitas and H. Videau, LC-TOOL-2003-010.
 
 \bibitem{Marlin} 
  F.~Gaede,
  Nucl.\ Instrum.\ Meth.\ A {\bf 559}, 177 (2006).

\bibitem{ILCSoft}  
http://ilcsoft.desy.de/portal

\bibitem{LCFIPlus} 
  T.~Suehara and T.~Tanabe,
  Nucl.\ Instrum.\ Meth.\ A {\bf 808}, 109 (2016).

\bibitem{PFA}
M. Thomson, Nucl.Instrum.Meth. A611 (2009) 25–40, arXiv:0907.3577.

\bibitem{KT_JET} 
  S.~Catani, Y.~L.~Dokshitzer, M.~H.~Seymour and B.~R.~Webber,
  Nucl.\ Phys.\ B {\bf 406}, 187 (1993).
  
\bibitem{FastJet} 
  M.~Cacciari, G.~P.~Salam and G.~Soyez,
  Eur.\ Phys.\ J.\ C {\bf 72}, 1896 (2012).

\bibitem{Durham} 
  Y.~L.~Dokshitzer, G.~D.~Leder, S.~Moretti and B.~R.~Webber,
  JHEP {\bf 9708}, 001 (1997).



\bibitem{ATLAS:2016eeo} 
  The ATLAS collaboration [ATLAS Collaboration],
  ATLAS-CONF-2016-059.

	
\bibitem{CMS:2016crm} 
  CMS Collaboration [CMS Collaboration],
  CMS-PAS-EXO-16-027.

	

\bibitem{Fujii:2016raq} 
  K.~Fujii {\it et al.} [LCC Physics Working Group Collaboration],
  arXiv:1607.03829 [hep-ph].


\end{thebibliography}
\end{document}